\documentclass[aps,pra,reprint,groupedaddress,showpacs]{revtex4-1}
\usepackage[english]{babel}
\usepackage{setspace}  
\usepackage{graphicx}
\usepackage{amssymb}
\usepackage{amscd}
\usepackage{rotating}
\usepackage{color}
\usepackage{epstopdf}

\newcommand{\sse}{\scriptstyle}
\newcommand{\sss}{\scriptscriptstyle}
\newcommand{\beq}{\begin{equation}}
\newcommand{\eeq}{\end{equation}}
\newcommand{\tra}[1]{\mathrm{Tr}\left\{#1\right\}}
\newcommand{\barr}{\begin{eqnarray}}
\newcommand{\earr}{\end{eqnarray}}
\newcommand{\stdif}[2]{\frac{\partial #1}{\partial #2}}
\newcommand{\tdif}[3]{\left(\frac{\partial #1}{\partial #2}\right)_{\sss #3}}
\newcommand{\tddif}[3]{\left(\frac{\partial^2 #1}{\partial #2^2}\right)_{#3}}
\newcommand{\tdiff}[3]{\frac{\partial^2 #1}{\partial #2\partial #3}}

\begin{document}

\title{Density-functional theory approach to the thermodynamics of the harmonically confined one-dimensional Hubbard model}

\author{V. L. Campo}
\email{vivaldo.leiria@gmail.com, vlcampo@df.ufscar.br}

\affiliation{Departamento de F\'isica, Universidade Federal de S\~ao Carlos, Rodovia Washington Lu\'is, km 235, Caixa Postal 676, 13565-905, S\~ao Carlos, SP, Brazil}

\date{\today}

\begin{abstract}
The thermodynamics of the inhomogeneous one-dimensional repulsive fermionic Hubbard model with 
parabolic confinement is studied by a density-functional theory approach, based on Mermin's 
generalization to finite temperatures. A local-density approximation (LDA), based on exact results 
for the homogeneous model, is used to approximate the correlation part in the Helmholtz free-energy, 
comprising the thermodynamic Bethe ansatz LDA (TBALDA). The general presentation of the method is given 
and some properties of the homogeneous model that are relevant to the DFT approach are analyzed. 
Extensive comparison between TBALDA and numerical exact diagonalization results for thermodynamic 
properties of small inhomogeneous chains is discussed. In the remaining, a classical thermodynamic 
treatment of the confined system is developed with the focus on global properties of large systems. 
A unusual behavior under isentropic expansion is found and discussed.
\end{abstract}
\pacs{03.75.Ss, 67.85.Lm, 05.30.Fk}
\maketitle

\section{Introduction}  

{\it Ab initio} calculations based on density-functional theory (DFT)~\cite{parrbook_kohnrev} 
have  become more and more popular in the last 30 years thanks to the continuous 
improvement on energy functionals, algorithms and available computational resources, with 
definitive impact on chemistry, physics and materials science. The original formulation 
by Hohenberg and Kohn~\cite{hk1964} was devised to the ground-state electronic density, 
however, applications in 
solid state physics and materials science usually demand that temperature be taken into account 
appropriately. Soon after the foundation of DFT by Hohenberg and Kohn, Mermin~\cite{mermin1965} 
generalized the formulation to the finite temperature case. The contribution of phonons to the 
thermodynamic properties of the system is absent from Mermin's formulation, but phonon 
spectra can be computed within DFT~\cite{baroniRMP2001} and their contribution 
to the physical properties can be incorporated~\cite{baroniRMP2001,qha2010}. Alternative 
approaches are under the umbrella of {\it ab initio} molecular dynamics~\cite{payneRMP92,tuckermanMDreview,abinitMDbook}. Accurate computation of the thermodynamic properties of materials at extreme conditions has had impact in the study of Earth's mantle~\cite{renata2010} and in the 
study of warm dens matter~\cite{desjarlais2002} for example. In applied physics, very complex 
systems, as many-components steels~\cite{steelRev}, have also been successfully approached by 
DFT.  Recent Quantum Monte Carlo numerical results for the temperature dependence of the 
exchange-correlation energy of the homogeneous electron gas~\cite{qmcT2013} and the proposed 
analytical fitting to this dependence~\cite{fitqmcT2014} will certainly allow improvement in the
free-energy functionals. 

In the last decade, DFT has been used to study model systems, such as the Hubbard 
model, the Heisenberg model and other models~\cite{reviewCapelleCampo}. The work for model 
systems has progressed similarly to the work in DFT to study realistic systems. Starting 
from ground-state DFT with simple functionals, there has been progress in functional development
and in the extension to treat time-dependent situations, but almost no work has been dedicated to 
treat systems at finite temperature. We mention here an application of DFT to 
the single impurity Kondo model~\cite{kurth_prl_2011} and the work by G. Xianlong {\it et al.}
~\cite{gao_temp13} on the one-dimensional fermionic repulsive Hubbard model, introducing a 
fitting to the correlation potential. 

The inhomogeneous fermionic Hubbard model can describe quite well systems of trapped 
ultracold atoms in optical lattices~\cite{ibloch_rmp2008}. Actually, one important motivation 
to build such ultracold atomic systems is to investigate experimentally the Hubbard 
model properties, given the relevance of the model to condensed matter physics and the fact that, 
despite 50 years of theoretical work, its physical properties remain only partially understood in 
two and three dimensions. In ultracold atomic systems, some external potential has to be present 
to confine the atoms, giving rise to inhomogeneous systems, promoting the approach to them 
by DFT, which will benefit from the fact that both the lattice and the confining potential are 
fixed, so that Mermin's formulation can be directly used here. 

In this work, the thermodynamics of the one-dimensional repulsive fermionic 
Hubbard model under parabolic confinement is obtained from a DFT approach 
according to Mermin's formulation. The correlation Helmholtz free-energy is 
approximated by the local-density approximation (LDA) based on exact results 
for the homogeneous model in the thermodynamic limit~\cite{takahashi1972,takahashibook,
takahashi2002,klumper1996,klumper1998,essler2005}, which is a natural extension to finite 
temperatures of the Bethe Ansatz LDA (BALDA) used to treat the inhomogeneous Hubbard model 
at $T=0$~K~\cite{reviewCapelleCampo}. For $T>0$~K, the approximation will be called here thermal
Bethe Ansatz LDA, or simply TBALDA. In the same way the homogeneous electron gas is the reference 
system to build the LDA used in {\it ab initio} calculations~\cite{parrbook_kohnrev}, the homogeneous
 Hubbard model is the reference system to build the BALDA and the TBALDA used here. The adoption of this reference system 
is crucial to the success of BALDA to treat the inhomogeneous Hubbard 
model~\cite{reviewCapelleCampo}. Within BALDA, not only the energy of the homogeneous system is 
exact, but also the energy gap in the charge sector of the half-filled homogeneous system is 
exactly taken into account~\cite{reviewCapelleCampo}. This energy gap is entirely due to the 
on-site interaction and makes the half-filled system insulating. In confined inhomogeneous systems, this energy gap is the reason for the coexistence of an insulating phase (corresponding to
a plateau with density equal to one in the density profile) and a surrounding metallic phase~\cite{rigol_dphases03,drummond_dphases05}.

%Aqui vou comentar sobre a rapidez de um calculo de DFT quando comparado com QMC ou DMRG, o que torna o metodo excelente para amplas varreduras de parametros, como as necessarias para uma detalhada analise das propriedades termodinamicas do sistema. Apesar do calculo via DFT nao ser exato, devido à aproximacao à energia de correlacao, em T=0K se verifica excelente concordancia com resultados exatos em ampla variedade de potenciais externos. (citar) Aqui mostramos comparacoes feitas com resultados exatos em uma cadeia de 9 sitios, os quais corroboram a boa qualidade do método de calculo via DFT.

%O interesse fundamental, no entanto, é o estudo de cadeias maiores, de tamanhos compatíveis com os encontrados experimentalmente, onde se espera que a LDA funcione bem melhor do que para cadeias pequenas.

%Despite the non-homogeneity of this confined system, one may ask whether 
%its thermodynamic behavior can be described in the same way we do for homogeneous 
%systems. It is shown that this is the case, after defining appropriately the system
%volume for this one-dimensional confined system. In this description, the non-homogeneity 
%is hidden but has direct consequences for the thermodynamic behavior. This is seen in the study
%of isentropic expansions. The experimental situation, where ultracold atomic systems are 
%thermally isolated and the confining potential can be tuned, is a natural motivation to study
%isentropic expansions.

Despite the inhomogeneity of this confined one-dimensional system, it is presented here a definition for its volume. What follows is that the thermodynamic behavior of the confined system can be described in the same way we do for homogeneous systems, having temperature, chemical potential and volume as independent variables. The inhomogeneity becomes apparently hidden, but has direct consequences for the thermodynamic behavior as it is shown in the present analysis of isentropic expansions. Considering isentropic expansions is naturally motivated by the experimental situation, where ultracold atomic systems are thermally isolated and the confining potential can be tuned. It is found here that in narrow
ranges of specific volume and temperature, the temperature in fact increases under isentropic expansion. This interesting behavior is linked to the changes that happen in the density profile as the confinement is relaxed and it is shown that it can be understood, in the framework of TBALDA, as a direct consequence 
of the peculiar density dependence of the correlation entropy in the homogeneous Hubbard model at low temperatures.

%Two usual assumptions are also made here. One is related to practical 
%aspects of computation and concerns the validity of using the grand-canonical ensemble. The 
%problem is that the cold atoms systems are not extremely large, when there would be no concern. 
%This has been discussed for other model systems and the results point to the validity of using
% the grand-canonical ensemble for systems of experimentally achievable 
%sizes~\cite{rigol_ensemble}. The second assumption concerns the more fundamental aspect of 
%thermalization and correctness of using standard statistical mechanics to describe this kind of 
%isolated confined cold atoms systems. Recent numerical studies of the time evolution of isolated 
%model systems have pointed out that, 
%after some thermalization time, their thermodynamic properties can be obtained by standard 
%statistical mechanics when the model is non-integrable. In our case, the presence of the 
%confining potential makes the model non-integrable and we assume thermalization accordingly. 

The work is organized as follows: section \ref{secmermin} presents the model Hamiltonian and 
Mermin's formulation to the model system is discussed in detail, section \ref{sectbalda} 
explains the thermodynamic Bethe-ansatz local-density approximation (TBALDA) used to the correlation 
Helmholtz free-energy, section \ref{sechom} discusses some properties of the homogeneous systems 
that are relevant to the DFT approach, section \ref{secshort} presents an extensive comparison 
between DFT results based on TBALDA and exact results for a short chain, where the approach can 
be seen to work well in spite of the smallness of the chain. In section \ref{secthermo} the 
thermodynamics of the harmonically confined Hubbard model is obtained from DFT calculations and 
a detailed study of the behavior of the system under isentropic expansion is presented. The 
conclusion follows in section \ref{secconclusion}.

\section{Background on Density-Functional Theory: Mermin's Theorem and Kohn-Sham scheme}
 \label{secmermin}
 
Consider a system described by the one-dimensional fermionic Hubbard model in the presence of an external potential, at fixed temperature ($T$) and chemical potential ($\mu$). The system hamiltonian is given by
\barr
\hat{H} &=& \hat{K} + \hat{U} + \sum_j v_j \hat{n}_j, \label{eqham} \\
\hat{K} &=& -t\sum_{j,\sigma} \left(c^\dagger_{j,\sigma} c_{j+1,\sigma} + \mathrm{h.c.}\right) \label{eqhamK}\\
\hat{U} &=& U\sum_j\hat{n}_{j,\uparrow}\hat{n}_{j,\downarrow}.\label{eqhamU}
% - \frac{1}{2}\right)\left(\hat{n}_{j,\downarrow} - \frac{1}{2}\right).
\earr
$c^\dagger_{j,\sigma}$ creates a fermion at site $j$ with spin $z$-component $\sigma$($=\pm 1/2$), $\hat{n}_{j,\sigma} = c^\dagger_{j,\sigma}c_{j,\sigma}$ is the spin-resolved site occupation operator at site $j$ and $\hat{n}_j = \hat{n}_{j,\uparrow} + \hat{n}_{h,\downarrow}$ is the total site occupation operator at site $j$. $\hat{K}$ represents the kinetic energy in this tight-binding model with hopping integral $t$. $\hat{U}$ represents the interaction energy operator, which increases by $U$ the energy of any doubly-occupied site. The external potential has amplitude $v_j$ at site $j$ and couples directly to the density. Hereafter we will express any energy in units of the hopping integral $t$.
% It would be simpler to consider for the interaction energy just $U\sum_j \hat{n}_{j,\uparrow}\hat{n}_{j,\downarrow}$, but the additional terms 
%while the third term was included to make $\hat{H}^{\mathrm{hom}}$ symmetric under a particle-hole transformation. Its inclusion is a matter of choice, having a trivial 
%effect on the chemical potential, but not changing the physics since the hamiltonian commutes with
%the particle number operator. The second and third terms will be refered to as % $\hat{U}$, so that 

Mermin's theorem~\cite{mermin1965} is the generalization of Hohenberg and Kohn theorem~\cite{hk1964,parrbook_kohnrev} to the case of fixed temperature and chemical potential, establishing the one-to-one correspondence between the external potential and the electronic density. Both theorems were originally established for continuous electronic systems with Coulomb interaction in mind. However, their generalization to discrete systems as the Hubbard model does not pose any difficulty. In this case, there is a one-to-one correspondence between the external potential set $\{v_j\}$ and the equilibrium site-occupations set $\{n_j\}$. For simplicity and to reinforce the parallelism with the continuous case, we will refer to site-occupations as densities. 

Let us briefly remind the reader of the main aspects of Mermin's theorem~\cite{mermin1965}. The thermodynamic properties of the system described by the hamiltonian in Eq.~(\ref{eqham}) are determined by the equilibrium density-matrix
\beq
\hat{\rho}_{eq} = \frac{e^{-\beta(\hat{H} - \mu \hat{N})}}{\tra{e^{-\beta(\hat{H} - \mu \hat{N})}}},
\eeq
which is the density-matrix that minimizes the functional
\beq
\Omega[\hat{\rho}] = \tra{\hat{\rho}\left(\hat{H} + k_{\sss B}T\ln(\hat{\rho}) - \mu \hat{N}\right)}.
\eeq
At equilibrium, $\Omega[\hat{\rho}_{eq}]$ is nothing else than the grand canonical potential,
\beq
\Omega[\rho_{eq}] = E - TS - \mu N = -k_{\sss B}T \ln(Z_G),\label{omegaofZG}
\eeq
where $E$ is the internal energy, $S$ is the entropy, $N$ is the number of
particles and 
\beq
Z_G = \tra{e^{-\beta(\hat{H} - \mu \hat{N})}} \label{ZGdef} 
\eeq
is the grand partition function.

From the minimizing property of $\hat{\rho}_{eq}$, Mermin proved the
one-to-one correspondence between the external potential and the equilibrium density in the same way Hohenberg and Kohn, based on the minimizing property of the ground-state wave function, proved such correspondence for an isolated system at $T=0$.

Since the density determines the external potential and the external potential determines $\hat{\rho}_{eq}$, we can introduce the universal density functional
\beq
{\cal F}[n](\mu,T) = \tra{\hat{\rho}_{eq}[n]\left(\hat{K} + \hat{U} + k_{\sss B}T\ln\left(\hat{\rho}_{eq}[n]\right) \right) },\label{eqFuniv}
\eeq
with no explicit reference to the external potential, which is itself a functional of the density. Following Mermin, for fixed $T$ and $\mu$ and for a given external potential, we define the density functional
\beq
\Omega_v[n](\mu,T,U) \equiv {\cal F}[n](\mu,T,U) - \mu N + \sum_j v_j n_j \label{eqOmega}
\eeq
and note that this functional is minimized by the equilibrium density 
\beq
n^{eq}_j = \tra{\hat{\rho}_{eq} \hat{n}_j},
\eeq
opening the door to make a variational approach to determine the equilibrium
density along the lattice as well as the thermodynamic properties of the system.
The equilibrium density is the minimum point of $\Omega_v[n](T,\mu,U)$, therefore
\beq
\stdif{\cal F}{n_j}[n_{eq}] + v_j - \mu = 0,\:\:\:\:\mathrm{any}\:\:j.\label{pmin}
\eeq
The main difficult is that the universal functional ${\cal F}[n](\mu,T)$ is not 
exactly known and some approximation to it is required to move on. Before considering the approximation to be used, let us make some remarks about the thermodynamics quantities of interest here.

For the equilibrium density, the functional $\Omega_v[n_{eq}]$ gives the grand canonical potential and $\Omega_v[n_{eq}] + \mu N$ gives the Helmholtz free energy. Two other quantities of interest are the entropy and the number of doubly occupied sites
\beq
N_D = \tra{\hat{\rho}_{eq}\sum_j \hat{n}_{j,\uparrow}\hat{n}_{j,\downarrow}}
= \tra{\hat{\rho}_{eq} \frac{\hat{U}}{U}}.\label{nd1}
\eeq

The entropy is given by $-\tdif{\Omega}{T}{\mu,T,U}$. When differentiating Eq.~(\ref{eqOmega}) with respect to the temperature, it is important to take into account that the site-occupations depend on the temperature, so
\barr
-S(\mu,T,U) &=&\sum_j\left[\stdif{\cal F}{n_j}[n_{eq}](\mu,T,U) + v_j - \mu \right]\tdif{n_j}{T}{}\nonumber\\ 
&+& \stdif{\cal F}{T}[n_{eq}](\mu,T,U),
\earr
 and from Eq.~(\ref{pmin}), we have
\beq
-S(\mu,T,U) = \stdif{\cal F}{T}[n_{eq}](\mu,T,U).\label{entropy1}
\eeq
Here, we have energies in units of $t$, temperature in units of $t/k_{\sss B}$ and entropy in units of $k_{\sss B}$.

For the number of doubly occupied sites, considering eigenstates of the hamiltonian to compute the trace in (\ref{nd1}) and the Hellmann-Feynman theorem, it is straightforward to get
\beq
N_D = \tdif{\Omega}{U}{\mu,T}.
\eeq
Differentiating Eq.~(\ref{eqOmega}) with respect to $U$ at the equilibrium density, we find
\beq
N_D = \stdif{\cal F}{U}[n_{eq}](\mu,T,U).
\eeq

According to the Kohn-Sham scheme, an auxiliar system of non-interacting fermions is considered with the requirement that its density be exactly the same as the density in the interacting system. This auxiliary non-interacting system is usually called Khon-Sham system. The universal functional for the interacting system will be decomposed as follows:
\barr
{\cal F}[n](\mu,T,U) &\equiv& {\cal F}_0[n](\mu,T) + E_{\mathrm{Hartree}}[n](U) \nonumber\\
&+&  {\cal F}_{c}[n](\mu,T,U). \label{eqFunivparts}
\earr
The first term on the right-hand-side of (\ref{eqFunivparts}) is the universal functional for
the non-interacting system ($U=0$), ${\cal F}_0[n](\mu,T) = F[n](T,\mu,0)$. The second term is the Hartree energy, here defined as
\beq
%E_{\mathrm{Hartree}}[n](U) = \frac{U}{4}\sum_j n_j^2 - \frac{U}{2}\sum_j n_j,\label{eqHartdef}
E_{\mathrm{Hartree}}[n](U) = \frac{U}{4}\sum_j n_j^2,\label{eqHartdef}
\eeq
corresponding to the first-order (in $U$) approximation to the expected value of $\hat{U}$. The third term, ${\cal F}_c[n](\mu,T,U)$, is the correlation energy functional, which encompass all the subtle many-body features of the universal functional. 

It is important to note that the correlation functional does not come only from the difference between the Hartree energy and the interacting energy $\langle \hat{U} \rangle$, but has a contribution due to the difference between the kinetic energy of the interacting system and the kinetc energy of the non-interacting system, as well as a contribution due to the difference between the entropies of these systems. In traditional DFT, the many-body term is referred to as exchange-correlation functional, but because the Hubbard model with one level per site does not have exchange energy, we use only correlation here.

With the decomposition in Eq.~(\ref{eqFunivparts}), the functional to be minimized (Eq.~(\ref{eqOmega})) is written as
\barr
&&\Omega_v[n](\mu,T,U) = {\cal F}_0[n](\mu,T) + \frac{U}{4}\sum_j n_j^2 \nonumber\\ 
&+& {\cal F}_c[n](\mu,T,U) + \sum_j v_j n_j - \mu N. \label{omega_veq}
\earr
At the exact density, we have $\frac{\partial \Omega_v[n](\mu,T,U)}{\partial n_j} = 0$ for any $j$, which implies
\barr
\frac{\partial {\cal F}_0[n](\mu,T)}{\partial n_j} + \frac{U}{2}n_j + \frac{\partial {\cal F}_c[n](\mu,T,U)}{\partial n_j} + v_j - \mu = 0. \label{eqKS1}\nonumber\\
\earr 
Defining the correlation potential by
\beq
v^c_j = \frac{\partial {\cal F}_c[n](\mu,T,U)}{\partial n_j},\label{eqvC}
\eeq
and the Kohn-Sham potential by
\beq
v^{KS}_j = v_j + \frac{U}{2}n_j + v^c_j, \label{eqKSpot}
\eeq
Eq.~(\ref{eqKS1}) becomes equivalent to
\beq
\frac{\partial {\cal F}_0[n](\mu,T)}{\partial n_j} + v^{KS}_j - \mu = 0, \label{eqKS2}
\eeq
which is exactly the equation we would have for a non-interacting system with
chemical potential $\mu$ in the presence of an external potential given by 
$v^{KS}$.

For a given Kohn-Sham potential $v^{KS}$, we can solve the non-interacting problem set by Eq.~(\ref{eqKS2}) determining the single-particle states and populating them according to the Fermi-Dirac distribution. In particular, the density for the Kohn-Sham system will be given by
\beq
n^{KS}_j = \sum_l \frac{1}{e^{\beta(\epsilon_l - \mu)} + 1}\langle\varphi_l | \hat{n}_j | \varphi_l\rangle \label{ksdenseq}
\eeq
where the sum runs over all the single-particle states $|\varphi_l\rangle$ with corresponding energies $\epsilon_l$. 

However, the Kohn-Sham potential is a functional of the density (see Eqs.~(\ref{eqvC}) and (\ref{eqKSpot})). For a given external potential, we minimize the functional $\Omega_v[n]$ in Eq.~(\ref{omega_veq}) starting with a trial density $n^0$. From $n^0$, we compute $v^{KS}$, solve the non-interacting problem and determine its density, $n^1$, by Eq.~(\ref{ksdenseq}). If $n^1 \neq n^0$, a new trial density based on $n^0$ and $n^1$ is suggested and the procedure is repeated until convergence. This is the self-consistent Kohn-Sham scheme.  The whole procedure would be exact if the correlation functional and its derivative were exactly known. With an approximate correlation functional, the Kohn-Sham scheme will give an approximation to the density, to the grand canonical potential and to other properties.

\section{TBALDA}
\label{sectbalda}

Here we will make the local-density approximation to the correlation energy. It is based on the exact Helmholtz free energy per site for the homogeneous Hubbard model with no external potential in the thermodynamic limit. This approximation will be referred to as thermodynamic Bethe ansatz local-density approximation, or more simply, as TBALDA. There are two main approaches to calculate the thermodynamic properties of the homogeneous Hubbard model. The first one, due to Takahashi~\cite{takahashi1972,takahashibook,takahashi2002}, is the one known as thermodynamic Bethe Ansatz. This approach leads to an infinite set of equations, which nonetheless can be treated nummerically to reasonable accuracy~\cite{takahashi2002}. The second approach is based on the quantum transfer matrix (QTM) method to deal with integrable systems and was formulated more recently~\cite{klumper1996,klumper1998,essler2005}. The QTM approach leads to a finite (and small) set of integral equations and was the approach followed by me to compute the properties of the homogeneous model, although the denomination used here suggests the opposite. The main reason to adopt this denomination is to make clear that the present work is a natural extension of previous work in density-functional theory applied to the Hubbard model~\cite{neemias2002,neemias2003,xianlong2006,reviewCapelleCampo}, where the acronym BALDA has become well established. 

The local-density approximation will be used in combination with the Kohn-Sham scheme~\cite{ks1965,parrbook_kohnrev} to approximate only the correlation part ${\cal F}_c[n](\mu,T,U)$ of the universal functional in Eq.~(\ref{eqFuniv}). We have
\beq
{\cal F}_c[n](\mu,T,U) \approx \sum_j f_c(n_j,T,U),\label{eqfclda}
\eeq
where $f_c(n_j,T,U)$ is the Helmholtz correlation energy per site of a homogeneous system in the thermodynamic limit with site ocupation $n_j$. According to Eq.~Eq.~(\ref{eqFunivparts}), $f_c$ is given by
\beq
f_c(n,T,U) = f(n,T,U) - f_0(n,T) - \frac{U}{4}n^2,\label{eqFC}
\eeq
where $f(n,T,U)$ and $f_0(n,T)$ are the Helmholtz free energies per site in the interaction and non-interacting systems, respectively. 
Therefore, within LDA, a generic site $j$ of the real inhomgeneous system contributes to ${\cal F}_c$ as it would contribute if it were part of an infinite homogeneous lattice with density equal to $n_j$ everywhere. 

On the right-hand-side of Eq.~(\ref{eqfclda}), the dependence on the chemical potential has disappeared and this deserves clarification. The point is that the reference infinite homogeneous system is considered with external potential {\it equal to zero}. If in the real system the site $j$ has occupation $n_j$, we need to consider a homogeneous system with site occupation equal to $n_j$ at all sites. In this homogeneous system, the chemical potential must be equal to $\mu(n_j,T,U) = \tdif{f}{n}{T,U}(n_j,T,U)$, where $f$ is the Helmholtz free energy per site for the homogeneous system. Alternatively, one could keep the chemical potential of the real system as an argument of $f_c$ in (\ref{eqfclda}), provided that the possibility of adding an uniform external potential $\tilde{v}$ to the homogeneous system was allowed. But this uniform external potential is equivalent to a shift in the chemical potential and when $\mu - \tilde{v} = \mu(n_j,T,U)$, we would have a system with site occupation $n_j$. 

\begin{figure}[t]
\centerline{\includegraphics[width=2.7in,keepaspectratio,angle=-90]{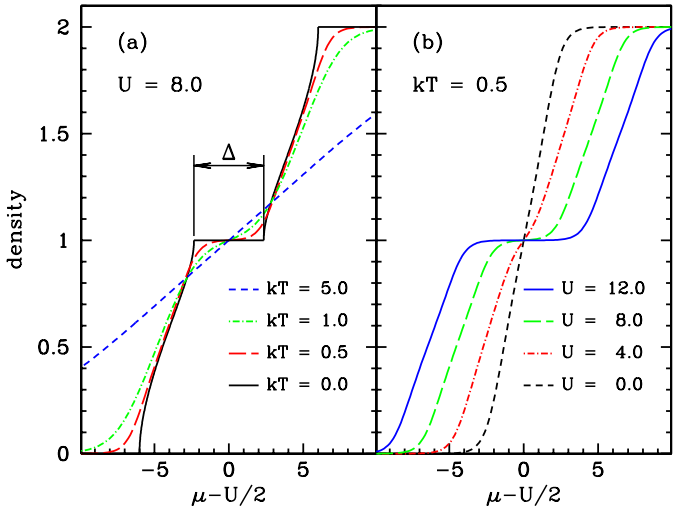}}
\caption{Density as a function of the chemical potential for the homogeneous one-dimensional Hubbard model. Energies are in untis of the hopping parameter $t$ (Eq.~(\ref{eqham})). The system is half-filled ($n=1$) when $\mu = U/2$. In this case, the system is insulating at $T=0$ and the density as a function of the chemical potential is constant and equal to one for $\frac{U}{2} - \frac{\Delta}{2} \leq \mu \leq \frac{U}{2} + \frac{\Delta}{2}$, where $\Delta$ is the energy gap in the charge sector. For $U=8.0$, $\Delta \approx 4.6795$. When the tempearature is small compared to $\Delta$, the above mentioned plateau is apparent.
\label{fig1m}}
\end{figure}

\begin{figure}[t]
\centerline{\includegraphics[width=2.6in,keepaspectratio,angle=-90]{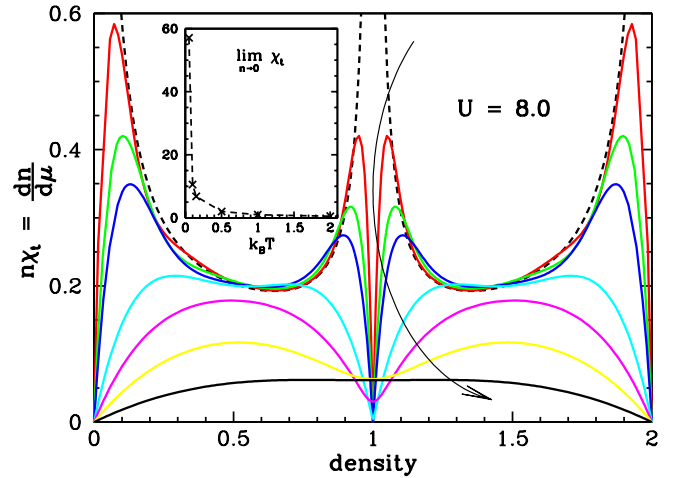}}
\caption{Density times the charge susceptibility (as defined in Eq.~(\ref{chieq})) as a function of the density for the homogeneous system with U=8.0 at several temperatures. The dashed line corresponds to $k_{\sss B}T=0$. The charge susceptibility diverges as the density approaches 0, 1 and 2, but if the density is exactly equal to one of these values, the charge susceptibility is equal to zero, as it is clear form Fig.~\ref{fig1m}. The other lines correspond to $k_{\sss B}T = 0.05$, $0.10$, $0.15$, $0.5$, $1.0$, $2.0$ and $5.0$ in the order the arrow crosses them. In the inset, we have the charge susceptibility in the limit of zero density. The finite value of this limit shows that $\tdif{n}{\mu}{T}$ behaves linearly with the density as $n\to 0$.
\label{fig2m}}
\end{figure}

The one-to-one correspondence between the chemical potential and the site occupation for the homogeneous system is illustrated in Fig.~\ref{fig1m} for different temperatures and $U$s. Additionally, the derivative of density with respect to the chemical potential as a function of the density is displayed in Fig.~\ref{fig2m} for several temperatures and $U=8$. 

Solving the integral equations coming from the QTM method, we determine $f(n,T,U)$. For $U=0$, it is simpler to make a direct calculation to get
\beq
f_0(n,T) = -\frac{2k_{\sss B}T}{\pi}\int_0^\pi \ln(1 + e^{-\beta(\epsilon_k - \mu_0)})~dk ~+~\mu_0 n. 
\eeq 
The single-particle energy is given by $\epsilon_k = -2t\cos(k)$ and the non-interacting chemical potential $\mu_0$ is an implicit function of the density $n$ through the relation
\beq
n = \frac{2}{\pi}\int_0^\pi~\frac{1}{e^{\beta(\epsilon_k - \mu_0)} + 1}~dk, \label{eqmu0}
\eeq 
where the Fermi-Dirac occupations for all extended single-particle states are added to build the total density.

%As mentioned before, the difference between the entropies 
%of the interacting and non-interacting systems is part of 
%the Helmholtz correlation energy ($f_c$). It is instructive to see the behavior% of 
%the entropy per site as a function of the density for the homogeneous Hubbard model. This illustrated in 
%Fig.~\ref{fig4m}.

From the approximate correlation functional, we obtain the aproximate correlation potential as follows:
\barr
v^c_j &=& \frac{\partial{\cal F}_c^{LDA}}{\partial n_j} = \frac{\partial f_c}{\partial n_j}(n_j,T,U)\nonumber\\
&=& \mu(n_j,T,U) - \mu_0(n_j,T) - \frac{U}{2}n_j, \label{eqVCpot}
\earr
where the chemical potential for the interacting homogeneous system is also obtained from integral equations coming from the QTM method, while for the non-interacting system, it is given by the implicit function in (\ref{eqmu0}). Accordingly, the Kohn-Sham potential at site $j$ (Eq.~(\ref{eqKSpot})) will be given by
\beq
v^{KS}_j = v_j + \mu(n_j,T,U) - \mu_0(n_j,T).\label{eqKSpotAlt}
\eeq

After convergence of the self-consistent Kohn-Sham scheme, we end up with an approximation to 
the density profile along the real interacting system. This density is the same as the density in
the auxiliary non-interacting system.  The grand canonical potential can be promptly determined from Eqs.~(\ref{eqOmega}), (\ref{eqFunivparts}), (\ref{eqHartdef}) and (\ref{eqfclda}),
\barr
&&\Omega_v[n_{eq}](\mu,T,U) =  {\cal F}_0[n_{eq}](\mu,T) + \sum_j \frac{U}{4}n_j^2\nonumber\\ 
&+& \sum_j f_c(n_j,T,U) + \sum_j (v_j - \mu) n_j,\label{eqOmegaKS}
\earr
where 
\beq
{\cal F}_0[n_{eq}](\mu,T) = \tra{\hat{\rho}_0\left(\hat{K} + k_{\sss B}T\ln(\hat{\rho}_0)\right)}
\eeq
and $\hat{\rho}_0$ is the equilibrium density-matrix for the non-interacting auxiliary system, whose single-particle energies are known and can be used to write
\barr
{\cal F}_0[n_{eq}](\mu,T) &=& -k_{\sss B}T\sum_l \ln\left(1 + e^{-(\epsilon_l - \mu)/k_{\sss B}T}\right)\nonumber\\
&-& \sum_j (v^{KS}_j - \mu) n_j.\label{eqF0}
\earr
In the first sum, $l$ runs over the single-particle states, while in the second sum, $j$ runs over the sites. Substituting Eq.~(\ref{eqF0}) back into Eq.~(\ref{eqOmegaKS}), we get 
\barr
&&\Omega_v(\mu,T,U) = -k_{\sss B}T\sum_l \ln\left(1 + e^{-(\epsilon_l - \mu)/k_{\sss B}T}\right)\nonumber\\
&+& \sum_j\left(v_j - v^{KS}_j + \frac{U}{4}n_j\right)n_j %\nonumber\\
+ \sum_j f_c(n_j,T,U). \label{omega_vtbalda}
\earr
%From Eqs.~(\ref{eqKSpot}), (\ref{eqFC}) and (\ref{eqVCpot}), we can alternatively write
%\barr
%\Omega_v(\mu,T,U) &=& -kT\sum_l \ln\left(1 + e^{-(\epsilon_l - \mu)/kT}\right)\nonumber\\
% &+& \sum_j \left(\Delta f_j - \Delta \mu_jn_j\right),
%\earr
%where
%\barr
%\Delta f_j &=& f(n_j,T,U) - f(n_j,T,0),\\
%\Delta \mu_j &=& \mu(n_j,T,U) - \mu(n_j,T,0).
%\earr

In the same way we related the entropy of the interacing system to the derivative of its universal functional with respect to the temperature (Eq.~(\ref{entropy1})), we have the following relation for the non-interacting Kohn-Sham system.
\beq
-S_{KS}(\mu,T) = \stdif{{\cal F}_0}{T}[n_{eq}](\mu,T).\label{entropy2}
\eeq
Therefore, the entropy can be computed differentiating Eq.~(\ref{eqFunivparts}). Within TBALDA for the correlation energy, we get
\barr
&&-S(\mu,T,U) =\sum_j\left[\stdif{{\cal F}_0}{n_j}[n_{eq}] + \frac{U}{2}n_j + v^c_j + v_j - \mu \right]\stdif{n_j}{T}\nonumber\\ 
&+& \stdif{{\cal F}_0}{T}[n_{eq}] - \sum_j\left(s(n,T,U) - s_0(n,T)\right).
\earr
From the Euler equation (\ref{eqKS2}), the term between square brackets in the first sum vanishes. From (\ref{entropy2}), we have
\beq
S(\mu,T,U) = S_{KS}(\mu,T) + \sum_j\left( s(n,T,U) - s_0(n,T)\right).\label{stbalda}
\eeq 
Analogously, for the number of doubly ocupied sites we have, within TBALDA,
\beq
N_D = \sum_j\stdif{f}{U}(n_j,T,U).
\eeq

For situations when the number of particles is fixed, we proceed with the canonical ensemble formalism. The adaptation of the previous formulation is straightforward. Instead of
the grand canonical potential, we have the Helmholtz free energy. The universal functional
is defined in the same way as done in Eq.~(\ref{eqFuniv}), but with the densities satisfying the constraint of fixed number of particles, $\sum_j n_j = N$. The functional to be minimized is
\beq
F_v[n](N,T,U) = {\cal F}[n](N,T,U) + \sum_j v_j n_j,
\eeq
which at the minimum will correspond to the equilibrium Helmholtz free energy. The universal functional can be decomposed as in Eq.~(\ref{eqFunivparts}), with the same expression for
the Hartree energy. The auxiliary Kohn-Sham system will have the effective potential at site $j$ given by the $v_j^{KS}$ in Eq.~(\ref{eqKSpot}). After solving the non-interacting
system, the site occupations have to be calculated in the canonical ensemble, which can
be considerably more difficult than in the grand canonical ensemble, due to the constraint on the number of particles. More specifically, if the systems are not large enough, the chemical potential is not well defined and the Fermi-Dirac distribution can not be used to populate the single-particle levels. For large enough systems, one could use a Fermi-Dirac distribution, setting up the chemical potential to have exactly $N$ particles. 
%As it will be seen in the following, even relatively small chains can be considered large enough for practical purposes.

Making the local-density approximation to the correlation functional,
the equations (\ref{eqfclda})--(\ref{eqKSpotAlt}) can be used. After convergence of the self-consistent Kohn-Sham scheme, one can calculate the Helmholtz free energy for the interacting system by
\barr
F_v(N,T,U) &=&  \tra{\hat{\rho}_0\left(\hat{K} + k_{\sss B}T\ln(\hat{\rho}_0)\right)}
+ \sum_j \frac{U}{4}n_j^2 \nonumber\\
&+& \sum_j f_c(n_j,T,U) + \sum_j v_jn_j.\label{eqFKS}
\earr
The first term on the right-hand side can be written in terms of the Helmholtz free energy of the non-interacting system $F_0(N,T) = -k_{\sss B}T\ln(Z_{0,N})$,
\beq
\tra{\hat{\rho}_0\left(\hat{K} + k_{\sss B}T\ln(\hat{\rho}_0)\right)} = F_0(N,T) - \sum_j v_j^{KS}n_j,
\eeq
where $Z_{0,N}$ is the partition function of the non-interacting system with $N$ particles. Therefore, we have
\barr
F_v(N,T,U) &=&  F_0(N,T) + \sum_j (v_j - v_j^{KS})n_j + \sum_j \frac{U}{4}n_j^2 \nonumber\\
&+& \sum_j f_c(n_j,T,U).\label{eqFKS2}
\earr

As a final remark in this section, it should be emphasized that the density-functional approach explained above and applied to the one-dimensional inhomogeneous Hubbard model could also be applied to the model in higher dimensions. The only practical difficulty would the absence of exact solution for the homogeneous reference system, making the construction of a reliable local-density approximation more difficult.

\section{The Homogeneous System}  
\label{sechom}

The homogeneous 1D Hubbard model, {\it i.e.}, the model in the thermodynamic limit 
with no external potential, can be treated exactly by the QTM method~\cite{klumper1996,klumper1998,essler2005}. This approach leads to a finite set of coupled integral equations for some auxiliary functions which in turn determine the grand canonical potential per site as a function of temperature, chemical potential and interaction strength $U$. By successive differentiation of the grand canonical potential with respect to these variables, all the thermodynamic quantities can be determined. %For convenience, the QTM integral equations as used in this work are presented in the appendix.

The physics of the homogeneous 1D Hubbard model at finite temperatures is well 
known~\cite{klumper1996,klumper1998,essler2005}, but in this section we will highlight 
aspects connected with the Helmholtz correlation energy per site ($f_c$ in 
Eq.~(\ref{eqFC})), which are relevant for TBALDA and have not been discussed in the 
literature so far.
% All the many-body subtleties in the density functional approach are incorporated in the correlation functional. Within LDA, it is of interest to study the behavior of $f_c$ and of the correlation potential $v^c = \frac{\partial f_c}{\partial n}$ introduced in Eq.~(\ref{eqVCpot}).

\begin{figure*}
\centerline{\includegraphics[width=5.4in,angle=-90]{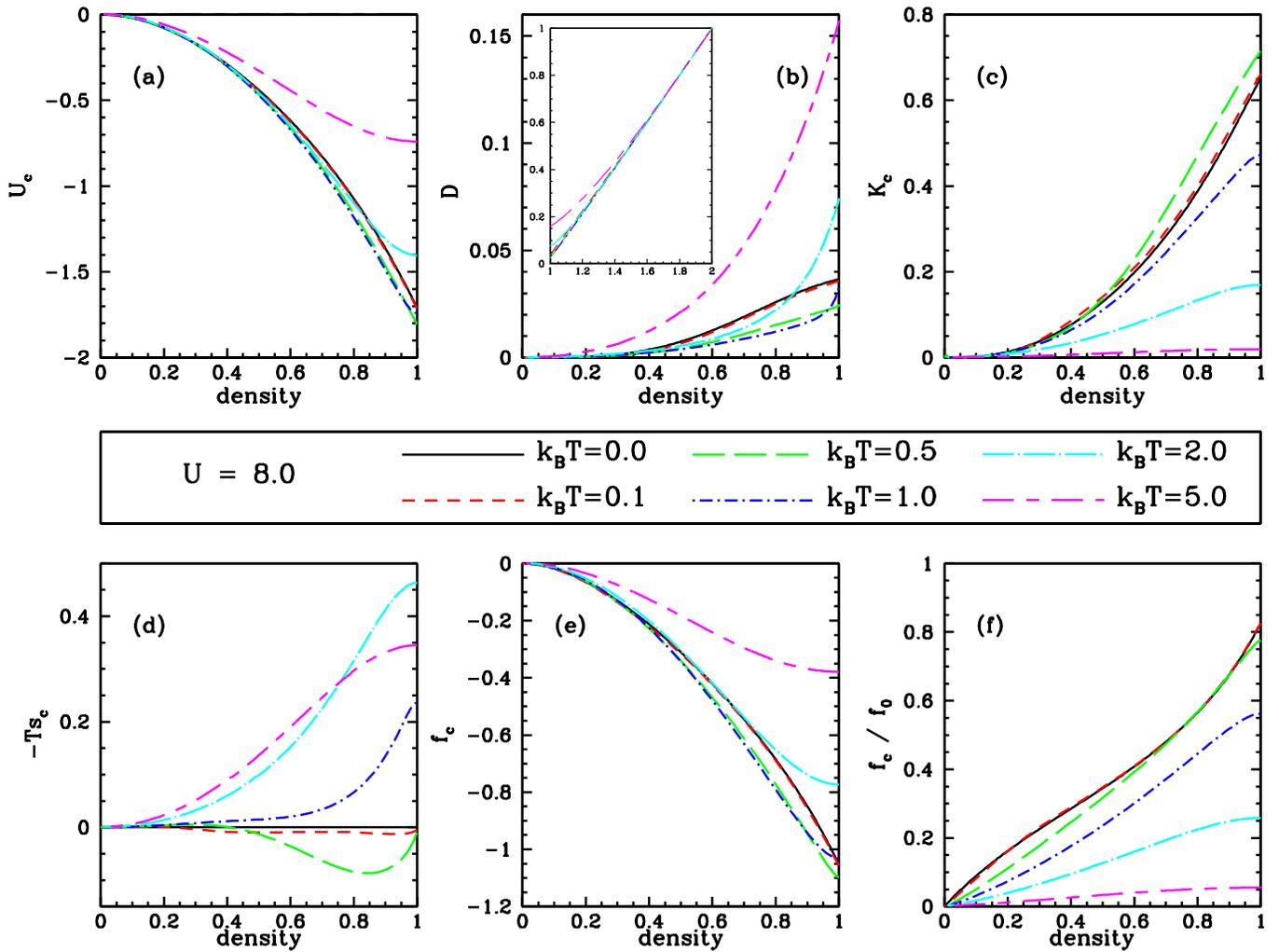}}
\caption{Decomposition of the Helmholtz correlation energy for U=8 as a function of the density at several temperatures. Energies are in units of $t$. The density interval was restricted to [0, 1] because the plotted functions are symmetric with respect to n = 1.  (a) The interaction component, $U_c = U(\langle \hat{n}_{j,\uparrow}\hat{n}_{j,\downarrow}\rangle
 - n^2/4)$. (b) The fraction of doubly-occupied sites $D = \langle \hat{n}_{j,\uparrow}\hat{n}_{j,\downarrow} \rangle$. This quantity is not symmetric with respect to $n=1$ and the inset shows its behavior for $n>1$, when it grows fast towards $1.0$ at $n=2$, when all the sites are totally occupied. 
(c) The kinetic energy component, $K_c$, is the difference between the kinetic energy per site in the interacting and non-interacting models. (d) The entropic component, $-Ts_c = -T(s - s_0)$, where $s$ and $s_0$ are
the entropies per site in the interacting and non-interacting models respectively. (e) The Helmholtz correlation energy obtained by adding up its three components, $f_c = K_c + U_c - Ts_c$.
(f) Ratio between the Helmholtz correlation energy and the Helmholtz free energy for the non-interacting system. For densities close to $n=1$ and temperatures smaller than the gap $\Delta$, the Helmholtz correlation energy is quite significant.
\label{fig3m}}
\end{figure*}

Considering interacting and non-interacting homogeneous systems at the same density $n$, 
we will decompose the Helmholtz correlation energy per site (Eq.~\ref{eqFC}) to
get a better understanding of its behavior. It will be written as $f_c = U_c + K_c -Ts_c$,
where $U_c$, $K_c$ and $-Ts_c$ are its interaction, kinetic and entropic components respectively. 
The interaction component is given by the difference between the interaction energy 
(Eq.~\ref{eqhamU}) per site and the first-order approximation to it,
\beq
U_c = \lim_{L\to \infty}\frac{\tra{\hat{\rho}\hat{U}}}{L} - \frac{Un^2}{4} = 
U\left[\langle\hat{n}_{j,\uparrow}\hat{n}_{j,\downarrow}\rangle - \frac{n^2}{4}\right] 
\eeq
where $L$ is the number of sites, taken to infinity in the thermodynamic limit with density $n$. The site $j$ in the final expression above is 
arbitrary, since the system is homogeneous in the thermodynamic limit. 
$\langle\hat{n}_{j,\uparrow}\hat{n}_{j,\downarrow}\rangle$ represents the fraction of
doubly-occupied sites in the system, which will be represented by $D$. Therefore, $U_c = U(D -n^2/4)$. From the Helmholtz free energy per site ($f$) as a function of $n$, $T$ and $U$, or from 
the grand canonical potential per site ($\omega = f - \mu n$) as a function of $\mu$, $T$ and $U$, we have
\beq
D = \tdif{f}{U}{T,n} = \tdif{\omega}{U}{T,\mu},
\eeq
which allows us to determine the interaction component $U_c$. In Fig.~\ref{fig3m}(a), we 
show $U_c$ as a function of the density at several temperatures for the case of $U=8$. 
Exploring a particle-hole transformation, it can be shown that $U_c(n) = U_c(2-n)$, so 
it is enough to consider densities from $0$ to $1$. The temperature dependence is weak 
while $k_{\sss B}T < \Delta$, where $\Delta$ is the gap in the spectrum of the charge sector 
for $n=1$ (when $U=8$, $\Delta \approx 4.6795$). As can be seen in Fig.~\ref{fig3m}(b), for densities $n \le 1$ and small temperatures, the average number of 
doubly-occupied sites is small, making the interacting energy small and the correlation 
component $U_c$ close to $-Un^2/4$. However, it is interesting to note that rising the temperature from $T=0$, the fraction of doubly-occupied sites, $D$, initially decreases with the temperature while we still have $k_{\sss B}T \ll \Delta$, and accordingly $U_c$ also 
initially decreases. This is due to the fact that at small temperatures there are more spin excitations (which do not demand double occupancy) than charge excitations (which tend to increase double occupancy). This behavior has been discussed in the context of thermometry and cooling strategies for ultra-cold trapped atomic systems made to simulate the Hubbard model~\cite{gorelik2012,vito2012}. As the temperature becomes close to or higher than 
$\Delta/k_{\sss B}$, the fraction of doubly-occupied sites grows, increasing $U_c$, whose magnitude decreases, however, approaching zero as $T \to \infty$ because the correlation disappears and $\langle\hat{n}_{j,\uparrow}\hat{n}_{j,\downarrow}\rangle \to \langle\hat{n}_{i,\uparrow}\rangle\langle\hat{n}_{j,\downarrow}\rangle = n^2/4$ in this limit.

The kinetic component $K_c$ is the difference between the kinetic energy per site in the interacting system and the kinetic energy per site in the non-interacting system. In the same way, the correlation entropy $s_c = s - s_0$ is the difference between the entropies per site in the interacting and non-interacting systems. Since the Helmholtz free energies and entropies can be obtained from QTM equations, the kinetic component is determined from 
$K_c = f_c - U_c + Ts_c$. In Fig.~\ref{fig3m}(c), we show $K_c$ as a function of the density at several temperatures for the case of $U=8$. It can be seen that $K_c > 0$ and, as $U_c$, it does not have a monotonic dependence with temperature. For small temperatures, $K_c$ is an increasing function of $T$, but as the temperature gets close to $\Delta$, this behavior is reversed and $K_c$ starts to decrease. In fact, this reversion starts at temperatures significantly smaller than $\Delta/k_{\sss B}$ for small densities. As $T\to \infty$, $K_c\to 0$ since correlation disappears at high temperatures.

\begin{figure}[t]
\centerline{\includegraphics[width=2.7in,keepaspectratio,angle=-90]{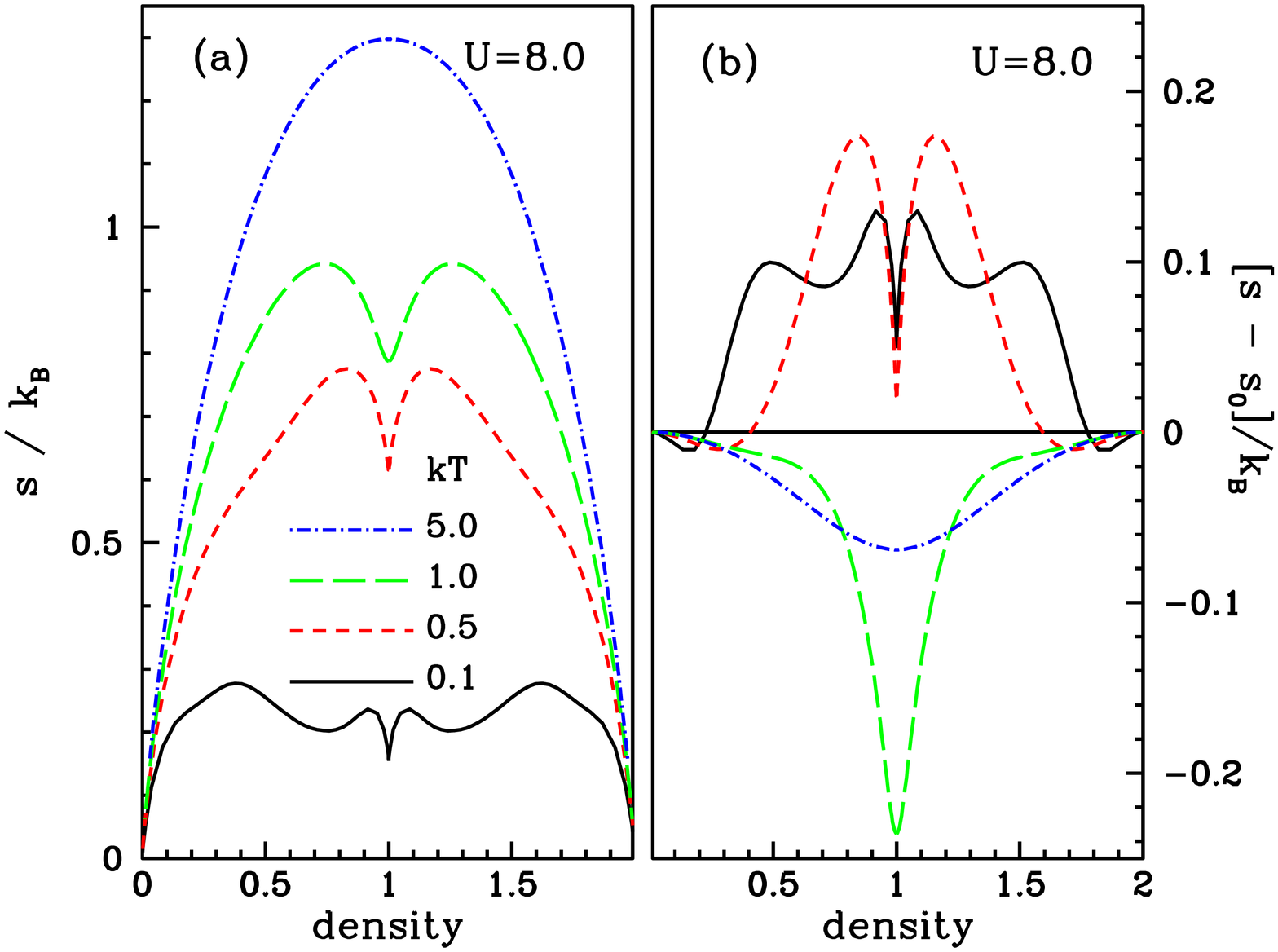}}
\caption{(a) Entropy per site in units of Boltzmann constant for the 1D homogeneous Hubbard model as a function of the density for $U=8t$ and several temperatures. Energies are in units of $t$. By a particle-hole transformation we see that the entropy per site must be symmetric with respect to $n=1$. For both $n=0$ and $n=2$, the entropy must be equal to zero. From the extrema towards half-filling, the way the entropy per site changes with the density depends strongly on the temperature. For high temperatures ($k_{\sss B}T > \Delta$), the entropy is maximum at $n=1$. For intermediate temperatures ($k_{\sss B}T < \Delta$), the entropy has a local minimum at $n=1$. At this point, the strong repulsion suppresses configurations with empty and doubly occupied sites giving rise to the local minimum of the entropy. For low temperatures, the entropy oscillates even more, but $n=1$ is still a local minimum.
 (b) Difference between the entropies per site of the interacting ($U=8t$) and non-interacting ($U=0$) systems.
%This difference is incorporated in the LDA treatment of the correlation functional presented in this work. 
For low temperatures, the interacting system has more entropy than the non-interacting one at almost all densities, while for higher temperatures we have the opposite behavior.
\label{fig4m}}
\end{figure}

In Fig.~\ref{fig3m}(d), we show $-Ts_c$, the entropic component of the Helmholtz free energy per site. In contrast to the other two components, the entropic one can be negative or positive (see also Fig.~\ref{fig4m}). Since $s_c = 0$ at $T=0$, $-Ts_c$ is very small at small temperatures, being initially negative. As the temperature is raised, the entropic component becomes positive and at intermediate temperatures (close to $\Delta/k_{\sss B}$), it is comparable to the kinetic component. As the temperature becomes larger compared to $\Delta/k_{\sss B}$, the correlation entropy $s_c$ becomes smaller and smaller, which will make the product $-Ts_c \to 0$ when $T\to \infty$. However, this limit is achieved much more slowly compared to the kinetic component and at high temperatures, the entropic and interacting components can have the same order of magnitude.

The whole Helmholtz correlation energy per site is displayed in Fig.~\ref{fig3m}(e), which is similar to Fig.~\ref{fig3m}(a), since the interaction component is the most important one. On general grounds, $f_c \le 0$ at any density and temperature. If $f_0$ corresponds to the Helmholtz free energy per site for the non-interacting system with the same density as the interacting one, we have
\beq
f = f_0 + U\frac{n^2}{4} + f_c.
\eeq
In Fig.~\ref{fig3m}(f), the ratio $f_c/f_0$ is displayed to better quantify how siginificant the contribution coming from correlation is. As expected, the precise comparison shows that correlation is specially important for densities close to $1$ and temperatures lower than $\Delta/k_{\sss B}$.

In Fig.~\ref{fig4m} the reader can see the behavior of the entropy per site (panel a) and of the correlation entropy per site (panel b) as functions of the density for the homogeneous system at several temperatures and $U=8$.

As discussed in the previous section, to the non-interacting or Kohn-Sham auxiliary system must be applied an effective potential $v^{KS}$ (Eqs.~(\ref{eqvC}) and (\ref{eqKSpot})) to keep its density equal to the density of the real interacting system. Within LDA, the correlation potential at each site (of a possibly inhomogeneous system) is given by the derivative $\tdif{f_c}{n}{T,U}$ evaluated at the site occupation, so some familiarity with the density dependence of $v_c = \tdif{f_c}{n}{T,U}$ is helpful to understand the behavior of the Kohn-Sham system.

\begin{figure}[t]
\centerline{\includegraphics[width=2.7in,keepaspectratio,angle=-90]{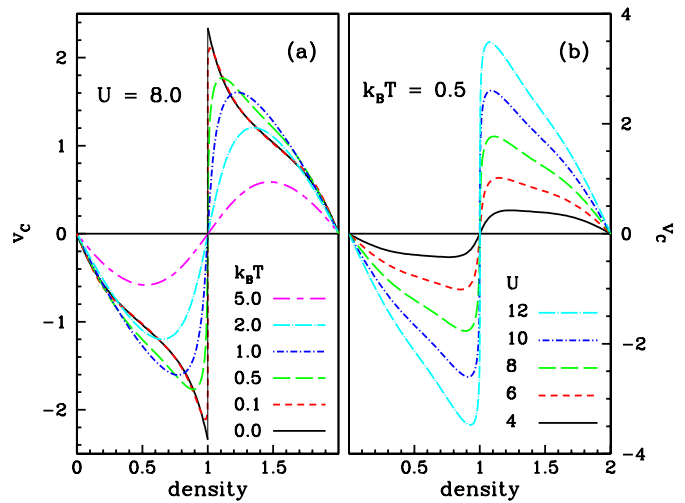}}
\caption{(a) Correlation potential ($v_c=df_c/dn$) as a function of density for $U=8$ and several temperatures. Energies are in units of $t$. At $T=0$, the correlation potential has a discontinuity at $n=1$ with a jump equal to the gap energy ($\Delta$). At finite temperatures, the discontinuity disappears. However, $v_c$ will change rapdily around $n=1$ while $k_{\sss B}T \ll \Delta$. (b) Correlation potential for $k_{\sss B}T = 0.5$ for several interaction strengths. As $U$ increases, the gap energy also increases, and the change of the correlation potential around $n=1$ becomes more pronounced.  
\label{fig5m}}
\end{figure}

From the QTM equations, we can get the density as a function of the chemical potential and the inverse of this function allows us to determine the correlation potential as indicated in Eq.~(\ref{eqVCpot}). As displayed in Fig.~\ref{fig5m}(a), at $T=0$ the correlation 
potential is discontinuous at $n=1$. The jump is equal to the gap $\Delta$~\cite{reviewCapelleCampo,neemias2002}. It is interesting to note that this gap is entirely due to correlation, since there is no gap in the Kohn-Sham spectrum of single-particle energies for a half-filled homogeneous system.
For finite temperatures, the discontinuity disappears, but $v_c$ changes fast around $n=1$ if $k_{\sss B}T \ll \Delta$. As the temperature is raised, the function $v_c$ becomes smoother, converging uniformly to zero in the limit $T\to \infty$. An abrupt change of the correlation potential around the density $n=1$ may sometimes make the convergence of the Kohn-Sham self-consistent loop very slow. This point has been discussed recently in Ref.~\cite{gao_temp13}.

Fig.~\ref{fig5m}(b) illustrates the behavior of $v_c$ for several values of interaction strength $U$ at the same temperature $k_{\sss B}T = 0.5$. For large $U$, we have $k_{\sss B}T \ll \Delta$, making $v_c$ be almost discontinuous at $n=1$. As $U$ decreases, so does $\Delta$, and the correlation potential becomes smoother and smother. For $U=0$, we would naturally have $v_c=0$ at any density.

\section{Comparison with exact diagonalization in a short chain}
\label{secshort}

\begin{figure*}[t]
\centerline{\includegraphics[width=5.0in,keepaspectratio,angle=-90]{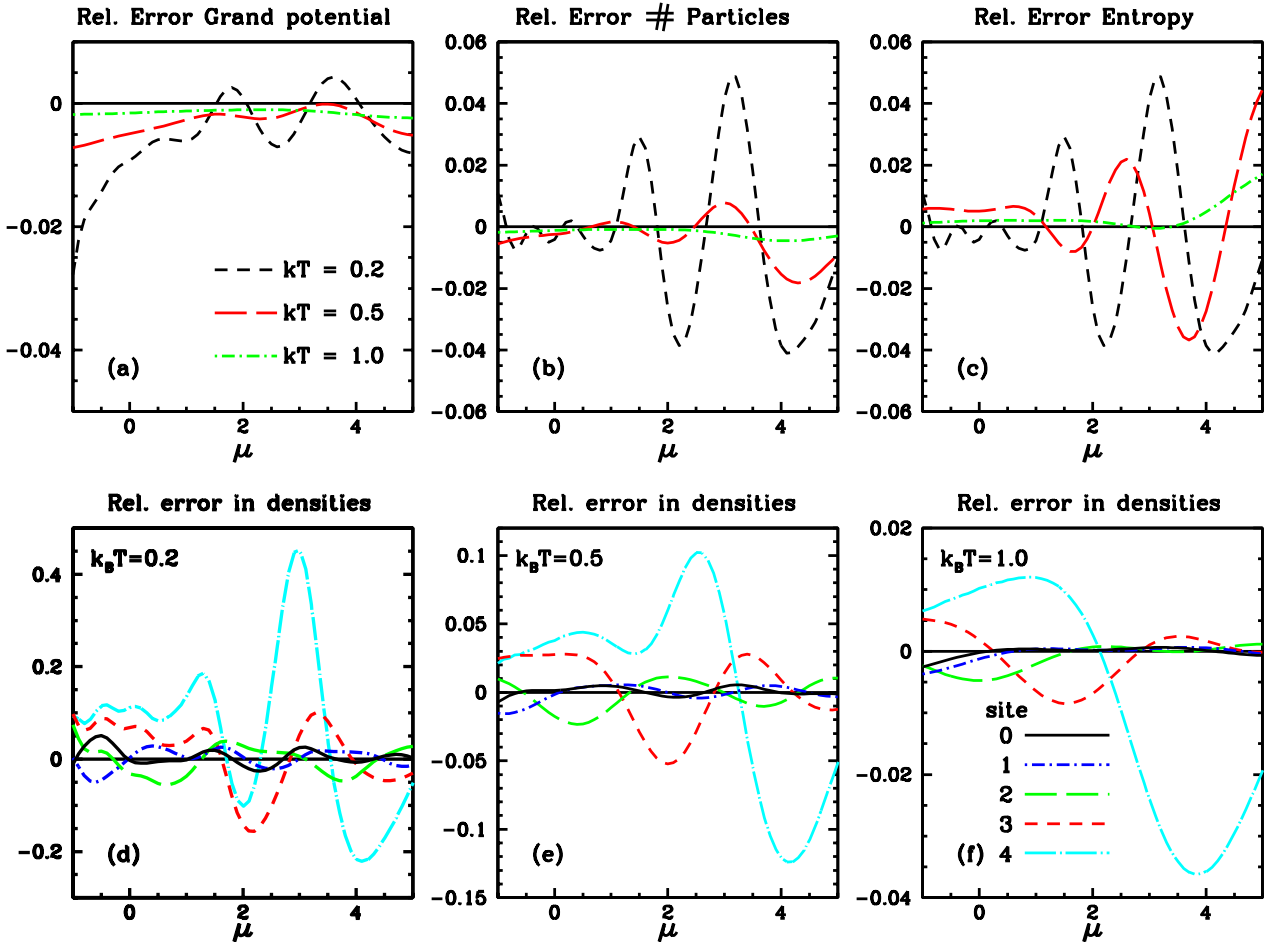}}
\caption{Figure to assess the performance of TBALDA for a Hubbard chain with 9 sites, submitted to an harmonic trap potential given by $V(i) = (i/2)^2$, with open boundary condition and $U=4$. Energies are in units of $t$. Relative errors are of form $(X-X_{\rm exact})/X_{\rm exact}$. %The large errors found in the density of the last site are mainly due the small density in this site. For a fixed chemical potential, lower the temperature, smaller that density. The large error in an almost empty site does not affect significantly the TBALDA prediction of the thermodynamic properties.
\label{fig6m}}
\end{figure*}

The formalism developed in section II can be used for any kind of external 
potential. Before comparing TBALDA and exact diagonalization in a short chain, it is important to explain how TBALDA was actually implemented. In this work, for given values of $U$, temperature and chemical potential, we solve the QTM equations for the homogeneous model to find the density, the correlation Helmholtz free-energy per site as well as its derivative with respect to the density (correlation potential) and the correlation entropy per site. Repeating the calculation for several values of chemical potential, we generate a fine mesh of densities in the interval $[0,2]$. Let us call this whole computation a mesh computation. For densities not presented in the mesh, a careful numerical interpolation is used to extract the quantities of interest keeping accuracy. Each mesh computation takes a few hours in a 2.4~GHz desktop computer. Given $U$ and $k_BT$, after the mesh computation, we can study within LDA the inhomogeneous Hubbard model for arbitrary external potential. However, for fixed external potential, the study of its properties as a function of temperature, requires a mesh computation for each temperature. It would be easier if we had accurate analytical fittings to the dependence on temperature and $U$ of the thermodynamic data from QTM equations. Although a very important first step in this direction has been given in Ref.~\cite{gao_temp13}, here we preferred to follow the more time demanding approach of doing a mesh computation for each temperature and $U$ to not introduce any additional error besides the already present error due to LDA.

 In this section, we consider a small Hubbard chain with 9 sites and
$U=4.0$, submited to the harmonic external potential $V_j = (j/2)^2$, $-4\le i \le 4$, with open boundary condition. The thermodynamic properties of this inhomogeneous system  were calculated from numerical exact diagonalization leading all eigenvalues and eigenvectors. We took into account the conservation of the number of particles and of the spin angular momentum ($S^2$ and $S_z$) to carry out the exact diagonalization in a few hours using a 2.4~GHz desktop computer. Since the number of eigenvalues grows exponentially with the chain size, going beyond nine sites would demand considerably greater computational resources than used here. Nine sites is already close to the maximum one could achieve computing all eigenvalues and eigenvectors, so it is a size neither too big, making exact calculations possible, nor too small, making the LDA aplicable. 

Having exact results, we can test the performance of TBALDA. A test with a small chain is a tough one because finite size effects and the spectrum discreteness are significant and can not be properly described by a local-density approximation based on results for the homogeneous system in the thermodynamic limit. If TBALDA performs relatively well in this test, we will have a strong indication that it will perform well enough in the large systems we are ultimately interested on. 
%%%%%%%%%%%%%%
All results discussed here are for the grand canonical ensemble. We analyzed the grand canonical potential, the entropy, the number of particles and the site densities in broad ranges of temperature and chemical potential. 
To better assesses the quality of TBALDA, we consider the relative errors. Once we have the density at the end of the self-consistent Kohn-Sham cycle, the
number of particles is trivially computed. The grand canonical potential is computed using Eq.~(\ref{omega_vtbalda}) and the entropy comes from Eq.~(\ref{stbalda}).

%\begin{table}[b]
%\caption{\label{tab1} Different ways to compute the entropy: 1- as in Eq.~(\ref{calcS}), 2- from $S = -\tdif{\Omega}{T}{\mu,U}$, 3- from $S = \sum_{i=1}^L s(T,n_i,U)$. Data for $k_{\sss B}T = 0.1$.}
%\begin{ruledtabular}
%\begin{tabular}{ccccc}
%$\mu$ & $S_1$ & $S_2$ & $S_3$ & $S_{exact}$ \\
%0.75 & 1.5084 & 1.5127 & 1.1279 & 1.3978 \\
%1.00 & 1.7221 & 1.7270 & 1.1660 & 1.4105 \\
%2.00 & 0.4070 & 0.4117 & 0.7075 & 0.4923 \\
%\end{tabular}
%\end{ruledtabular}
%\end{table}

Fig.~\ref{fig6m} illustrates the whole scenario. The first important feature is 
that for almost all the displayed curves, the relative errors are really 
small, less than 5\% in most of the points. The second important feature is 
that the relative errors increase as the temperature is lowered, which is 
natural because at low temperatures the quantum fluctuations are more 
prominent and it will be more difficult to capture the consequences of them 
using a LDA to treat this small system. 

From the extensive comparison between BALDA and QMC or DMRG results at $T=0$~\cite{xianlong2006,Lima2007557}, that usually indicates small errors in the ground-state energy and site occupations, we could expect small erros in the grand canonical potential, site occupations and number of particles at finite temperatures and it is exactly what we observe in Fig.~\ref{fig6m}.

The entropy tipically gives rise to a small contribution to the grand canonical potential
(through the $-TS$ term) and its determination requires an accurate description of the temperature dependence of the grand canonical potential. The derivative of our approximate grand canonical potential functional with respect to the temperature at its minimum is given by Eq.~(\ref{stbalda}). It is therefore unnecessary to make numerical derivatives to compute the TBALDA approximation to the entropy. In panels (b) and (c) of Fig.~\ref{fig6m}, we can see that the errors in the entropy and the errors in the number of particles are of same order of magnitude. 

In the second line of Fig.~\ref{fig6m}, we can see the relative errors in the approximated 
site-occupations. Due to the symmetry of the external potential, we only show 
data for the central site ($i=0$) and for the sites on the right of it ($i>0$). The general trend of larger errors at lower temperatures is evident. Due to 
the parabolic external potential, the site-occupation at the endings of the 
chain will be small. At low temperatures, this effect is stronger, once the 
particles do not gain thermal energy enough to reach the endings of the chain. 
Accordingly, we can understand the large relative errors in the density of site 4 and understand that the error decreases as we move to the center of the chain.  
Therefore we can conclude that TBALDA has been successful in this tough test 
of a small chain. One can treat larger systems, where typically the external 
potentials are  smoother than here, with confidence that the thermodynamic 
properties will be approximated within a few percent of accuracy. In the next 
section, we will consider the case usually found in atomic trapped systems 
with optical lattices, which is the main motivation for extending the BALDA 
approach to finite temperatures. 

\section{Thermodynamics of the harmonically confined one-dimensional Hubbard model}  
\label{secthermo}

\subsection{General aspects}

\begin{figure}[t]
\centerline{\includegraphics[width=2.6in,keepaspectratio,angle=-90]{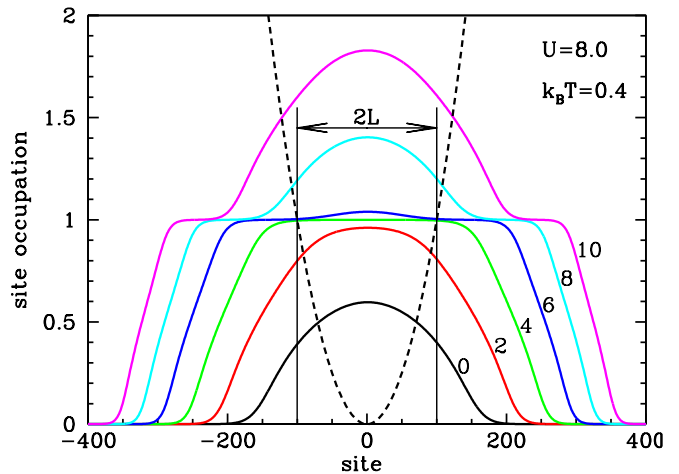}}
\caption{Density profiles for system ``volume'' $2L = 200$, $U=8.0$ and $k_{\sss B}T=0.4$ for different values of the chemical potential ($\mu = 0$, $2$, $4$, $6$, $8$ and $10$). The dashed line represents the parabolic potential $V(x) =(x/100)^2$. As the chemical potential increases, a significant fraction of the particles will be found outside the interval $[-L,L]$. As the temperature is much smaller than the gap ($k_{\sss B}T/\Delta \approx 0.85$), the central plateau around density equal to $1.0$ is visible for $\mu=4$, as well as lateral plateaus are also visible for larger values of the chemical potential. Energies are in units of $t$. 
\label{fig7m}}
\end{figure}

In this section, we consider the case of harmonic traps, whose 
external potential has the general form
\beq
v_j = (j/L)^2. 
\eeq  
Accordingly, $2L$ can be considered the typical size of the confined 
system. In what follows it will be helpful to refer to this size ($2L$) as 
the volume ($V$) of the system. Given the general familiarity with classical 
thermodynamics, this abuse of notation will make our treatment more appealing.
Fig.~\ref{fig7m} displays typical density profiles and makes clear that in a
harmonically confined system the number of particles outside its volume $2L$ 
can be quite significant. In spite of this fact and of the characteristic inhomgeneity of the system, we show below that its global thermodynamic description can be done in a way totally similar to the classical thermodynamics of homogeneous systems. 

\begin{figure}[t]
\centerline{\includegraphics[width=2.7in,keepaspectratio,angle=-90]{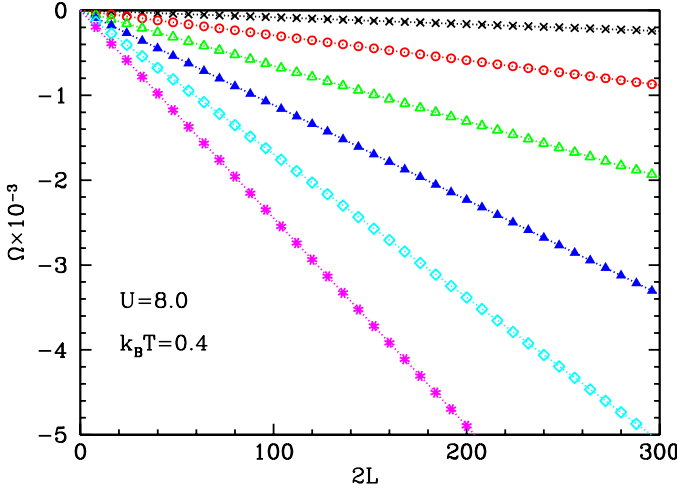}}
\caption{Grand canonical potential ($\Omega$) as a function of the volume ($V=2L$) at $k_{\sss B}T=0.4$ for different values of the chemical potential. Energies are in units of $t$. From top to bottom we have $\mu = 0$, $2$, $4$, $6$, $8$ and $10$. The linear behavior since small sizes is clear from the excellent distribution of the points over the linear fittings to them (dotted lines). This picture is only illustrative of this fact which was observed for all temperatures, chemical potentials and interaction strengths in broad ranges investigated. 
\label{fig8m}}
\end{figure}

Consider the behavior of usually extensive quantities, like internal energy, 
free energy, entropy and number of particles of usual systems with sharp boundaries as we increase the volume keeping temperature and chemical potential fixed. In the thermodynamic limit, when surface effects become negligible, any extensive quantity is proportional to the volume. Figure \ref{fig8m} illustrates what happens for the inhomogeneous harmonically confined Hubbard model. The linear dependence of the grand canonical potential (and any other extensive property)
on the volume is observed even for relatively small volumes, opening
the possibility for a classical thermodynamic treatment of this system.

The fact that a harmonically confined system has no surface, with the external potential growing slowly, suggests that the finite-size correction to the thermodynamic limit of the ratio between any extensive quantity and the volume $2L$ can be special. In fact, for the case of the ground-state energy, it was found~\cite{physicaB} that $E/(2L) = \epsilon_\infty + 1/(2L/\xi)^\gamma$, with the exponent $\gamma  > 1$ (around $1.4$ for $U=2.0$). This exponent larger than the common value of $1$ makes the thermodynamic limit $\epsilon_\infty$ rapidly achievable and explains the results also found for finite temperatures in Fig.~\ref{fig8m}.
   
Having the chemical potential, the temperature, the interaction strength ($U$) and the volume as independent variables, the grand canonical potential, $\Omega(\mu,T,U,V)$, will give the complete thermodynamic
description of the system. In particular, we have
\beq
d\Omega = -S dT -Nd\mu + N_DdU - pdV, \label{dOmega}
\eeq
where $S$ is the entropy, $N$ is the number of particles, $N_D$ is the number of doubly occupied sites and $p$ is the pressure. But what does the pressure mean in this harmonically confined Hubbard model? To answer this, it is enough to work out from Eq.~(\ref{omegaofZG}) relating the grand canonical potential to the
grand partition function $Z_G = \tra{e^{\beta\left(\hat{H} - \mu\hat{N}\right)}}$. We
have
\barr
&&\tdif{\Omega}{V}{\mu,T,U} = \frac{1}{Z_G} \sum_R \frac{\partial}{\partial V}\left(E_R - \mu N_R\right)e^{-\beta(E_R - \mu N_R)}\nonumber\\
&=& \frac{1}{Z_G} \sum_R \frac{\partial}{\partial V}\langle R | \hat{H} - \mu \hat{N} | R \rangle\: e^{-\beta(E_R - \mu N_R)}\nonumber\\
&=& \frac{1}{Z_G} \sum_R \langle R | \frac{\partial\hat{H}}{\partial V} | R \rangle \:e^{-\beta(E_R - \mu N_R)}= -\frac{\langle\hat{V}_{\rm ext}\rangle}{L}. 
\earr
where $\hat{V}_{ext} = \sum_j \left(\frac{j}{L}\right)^2\:\hat{n}_j$ is the external potential operator due to the harmonic trap. Therefore, the trap pressure is given by
\beq
p = \frac{\langle \hat{V}_{\rm ext}\rangle}{L} =
\frac{2\langle \hat{V}_{\rm ext}\rangle}{V}.\label{trapp}
\eeq
The proportionality between $\Omega$ and volume in the harmonically confined 
system seen in Fig.~\ref{fig8m} can be mathematically expressed by
\beq
\Omega(\mu,T,U,\lambda V) = \lambda\Omega(\mu,T,U,V)
\eeq  
for real $\lambda$. This immediatly gives us the familiar Euler relation 
\beq
\Omega(\mu,T,U,V) = -p(\mu,T,U)V,
\eeq 
where the trap pressure is a function of temperature, chemical potential and $U$ only. 
%At this point, it is rewarding to see that the trap pressure computed directly 
%from Eq.~(\ref{trapp}) is in fact equal to $-\Omega/V$ (within numerical 
%precision) as displayed in Fig.~\ref{fig9}. 

%\begin{figure}[t]
%\centerline{\includegraphics[width=2.7in,keepaspectratio,angle=-90]{./figures/fig9_ks.ps}}
%\caption{Ratio between the grand canonical potential ($\Omega$) and the product $pV = 2pL$
%as a function of the chemical potential for temperatures ranging from $kT=0.1$ to $kT=1.0$ with step $0.01$. Energies are in units of $t$. Within numerical accuracy, this ratio is equal to -1 as should be expected as a consequence of the linear behavior in Fig.~\ref{fig8m}. 
%\label{fig9}}
%\end{figure}

Any other thermodynamic function can be computed from derivatives of $p(\mu,T,U)$. 
 In particular, the first derivatives give the particle number, the entropy and the number of doubly occupied sites,
\barr
N(\mu,T,U,V) &=& \tdif{p}{\mu}{T,U}\:V, \label{NmuT}\\
S(\mu,T,U,V) &=& \tdif{p}{T}{\mu,U}\:V, \label{SmuT}\\
N_D(\mu,T,U,V) &=& -\tdif{p}{U}{\mu,T}\:V, \label{NDmuT}
\earr
so
\beq
dp = \frac{N}{V}d\mu + \frac{S}{V}dT - \frac{N_D}{V}dU,
\eeq
that is the Gibbs-Duhem relation between the differentials of the four intrinsically intensive quantities for this kind of thermodynamic system.  In the following, we will assume constant $U$ and will omit this variable for simplicity.
The $U$-dependence is a very important feature of atomic traps and will be studied more carefully along this line in a future work. 

Let us introduce some thermodynamic coefficients obtained from the second derivatives of the trap pressure. We have the isothermal charge susceptibility, 
 \beq
\chi_t = \frac{1}{N}\tdif{N}{\mu}{T,V} = \frac{V}{N}\tddif{p}{\mu}{T},\label{chieq} 
\eeq
the specific heat at constant $\mu$ and V,
\beq
c_{\mu,v} = \frac{T}{N}\tdif{S}{T}{\mu,V} = \frac{TV}{N}\tddif{p}{T}{\mu},\label{sheq}
\eeq
and the thermal particle-increment coefficient,
\beq
\nu = \frac{1}{N}\tdif{N}{T}{\mu,V} = \frac{V}{N}\tdiff{p}{T}{\mu}. \label{mueq}
\eeq
Note the Maxwell relation $\tdif{S}{\mu}{T,V} = \tdif{N}{T}{\mu,V} = \nu N$.
While $\chi_t$ and $c_{\mu,v}$ are never negative as a consequence of the second law of thermodynamics, the thermal particle-increment coefficient can assume positive and negative values.

With the above definitions we can write
\barr
dp &=& \frac{s}{v}dT + \frac{1}{v}d\mu, \label{dpeq}\\
dN &=& \nu NdT + \chi_t Nd\mu + \frac{1}{v} dV, \label{dNeq}\\
dS &=& \frac{c_{\mu,v}}{T}NdT + \nu N d\mu + \frac{s}{v}dV, \label{dSeq}
\earr
where $s=S/N$ is the entropy per particle and $v=V/N$ is the volume per particle.
From the differentials in Eqs.~(\ref{dpeq})--(\ref{dSeq}), any other thermodynamic coefficient can be written in terms of $\chi_t$, $c_{\mu,v}$ and $\nu$. Some of them are listed in Table \ref{tab1}.

\begin{table}[b]
\caption{\label{tab1} Common thermodynamic coefficients and their relation to those defined in Eqs.~(\ref{chieq})--(\ref{sheq}).}
\begin{ruledtabular}
\begin{tabular}{cc}
\begin{tabular}{c}\small adiabatic charge\\\small susceptibility\end{tabular} & $\chi_s {\sse=} \frac{1}{N}\tdif{N}{\mu}{S,V} {\sse =}\: \chi_t - \frac{\nu^2T}{c_{\mu,v}}$\\
\begin{tabular}{c}\small thermal expansion\\\small coefficient\end{tabular} & $\alpha {\sse =} \frac{1}{V}\tdif{V}{T}{p,N} {\sse =}\: s\chi_t - \nu$\\
\begin{tabular}{c}\small specific heat at\\\small constant volume\end{tabular} & $c_v {\sse =} \frac{T}{N}\tdif{S}{T}{V,N} {\sse =}\: c_{\mu,v} - \frac{\nu^2T}{\chi_t}$\\
\begin{tabular}{c}\small specific heat at\\\small constant pressure\end{tabular} & $c_p {\sse =} \frac{T}{N}\tdif{S}{T}{p,N} {\sse =}\: c_{\mu,v} + Ts(s\chi_t -2\nu)$\\
\begin{tabular}{c}\small isothermal\\\small compressibility\end{tabular} & $\kappa_t {\sse =} -\frac{1}{V}\tdif{V}{p}{T,N} {\sse =}\: v\chi_t$\\
\begin{tabular}{c}\small adiabatic\\\small compressibility\end{tabular} & $\kappa_s {\sse =} -\frac{1}{V}\tdif{V}{p}{S,N} {\sse =}\: \frac{c_{\mu,v}}{c_p}v\chi_s {\sse =}\: \frac{c_v}{c_p}\kappa_t$
\end{tabular}
\end{ruledtabular}
\end{table}
%
%1. comparar a) full KS,
%            b) local chemical potential
%
%                  b.1) de mu_i -> n_i -> LDA para gpot, entropy, D, etc
%
%                  b.2) de mu_i -> n_i -> 1 ciclo de KS -> T_s, S_s, etc -> LDA 
%                       para os termos de correlacao apenas.
%
%

\subsection{Isentropic expansion}
\label{secexpan}

\begin{figure}[t]
\centerline{\includegraphics[width=3.5in,keepaspectratio]{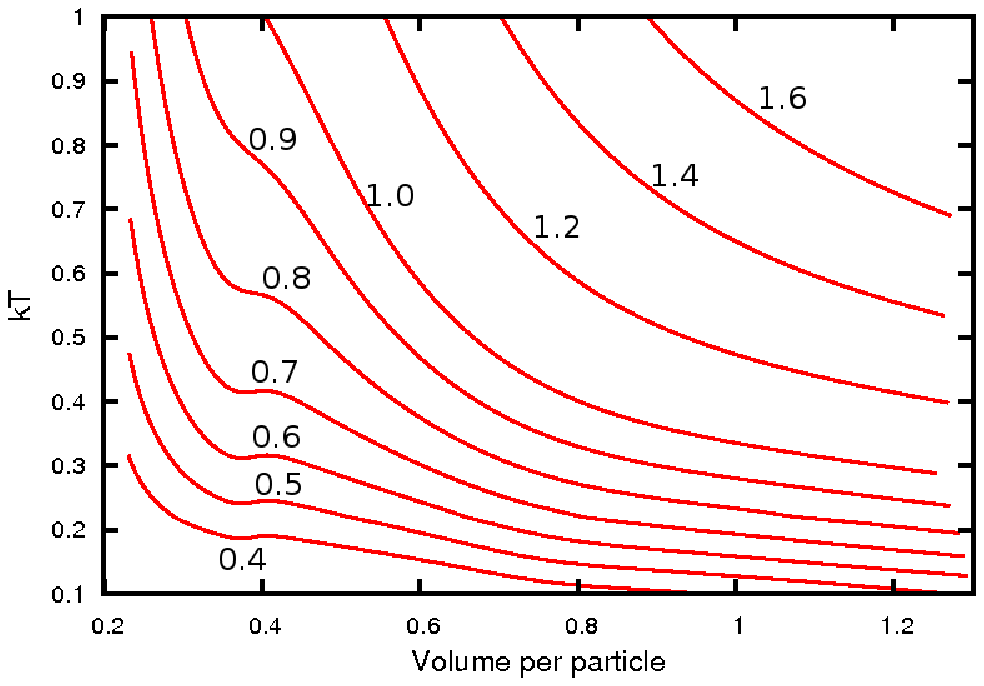}}
\centerline{\includegraphics[width=2.5in,keepaspectratio,angle=-90]{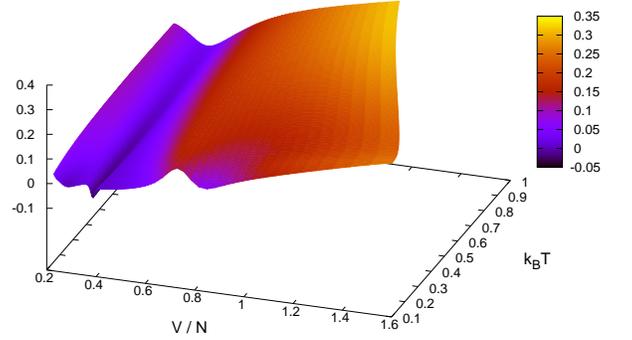}}
\caption{Top: temperature evolution in isentropic expansions of harmonically trapped systems. The values of entropy per particle in units of Boltzmann constant are indicated for each curve. Data are for $U=8.0$. For small values of entropy per particle ($\lesssim 0.75$), there are narrow intervals of volume where the temperture increases during the expansion. Bottom: thermal expansion coefficient ($\alpha$) as a function of volume per particle and temperature. At low temperatures and for $V/N$ in a small range below 0.4, $\alpha$ is negative. It is valid to note that fast variations of this coefficient are related to qualitative changes in the density profile. Other coefficients in table~\ref{tab1} also present this behavior.
\label{fig9m}}
\end{figure}

\begin{figure*}
\centerline{\includegraphics[width=4.5in,angle=-90]{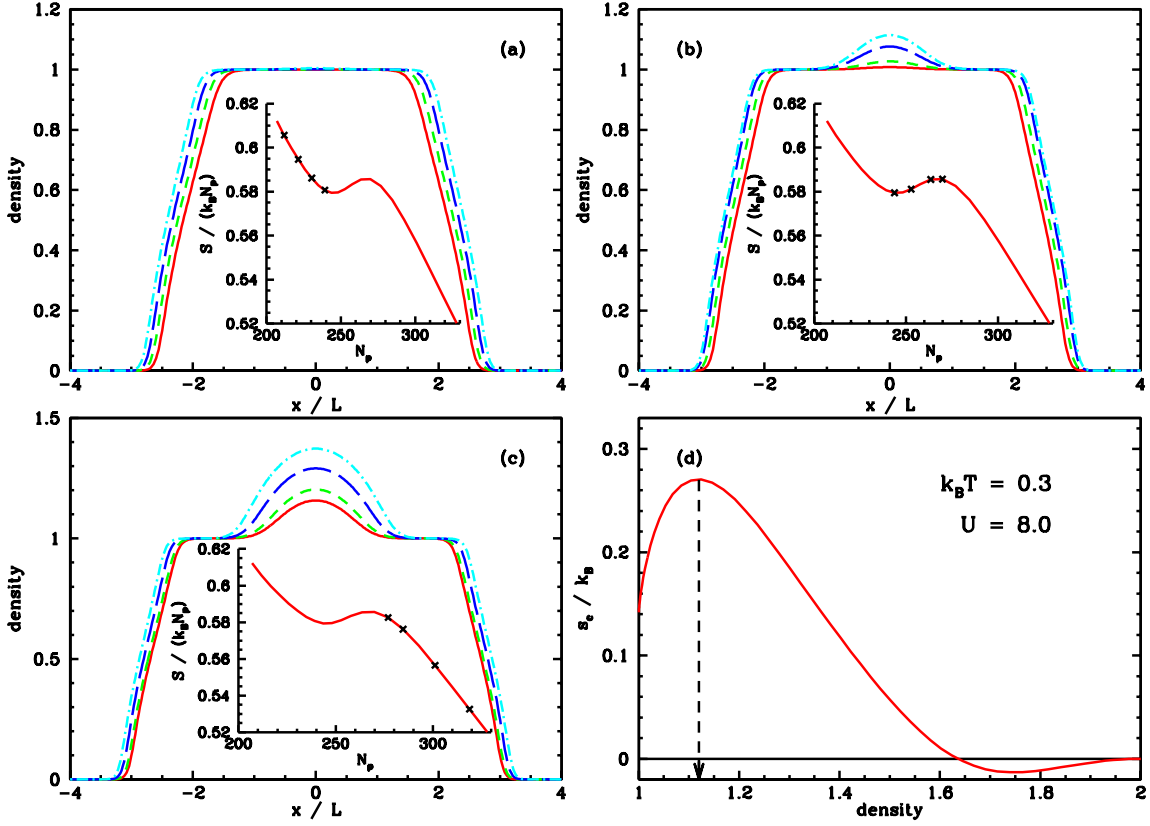}}
\caption{Relation between the density profiles and the non-monotonic behavior of the entropy per particle as a function of number of particles at constant temperature ($k_BT = 0.3$) and volume ($2L = 100$). Data for $U=8.0$. Panels (a), (b) and (c) show the density profile in the trap for different numbers of particles. The inset in each of those panels shows the dependence of entropy per particle on number of particles. The crosses correspond to the density profiles in the main plot. Panel (d) displays the correlation entropy per particle as a function of the density for the homogeneous model. The graph is symmetric around $n=1.0$ and only the curve for $n\ge 1$ is shown. It has a local minimum at $n=1.0$ and a close local maximum at $n=1.12$. Similar plots for other temperatures are shown in Fig.~\ref{fig4m}b.\label{fig10m}}
\end{figure*}

If the parabolic external potential has its curvature slowly changed to allow thermodynamic equilibrium, we have a thermodynamic transformation with constant particle number (for negligible loss of particles) and constant entropy (there is no heat transfer). The change in curvature represents a change in the volume ($2L$) as defined above. Figure \ref{fig9m} displays the temperature as a function of the volume per particle for several values of entropy per particle. The usual adiabatic cooling under expansion is observed almost everywhere. However, for a narrow interval of specific volumes around 0.4 sites per particle and specific entropy $\lesssim 0.75~k_B$, we see temperature increasing instead. Such behavior is unusual but it is thermodynamically possible because the thermal expansion coefficient ($\alpha$ in Table~\ref{tab1}) can be negative (as it happens, for example, with water below 4~$^\mathrm{o}$C at 1~atm) and
\beq
\tdif{T}{v}{s} = -\frac{\alpha T}{c_v \kappa_t}.
\eeq
The behavior of the thermal expansion coefficient is displayed in Fig.~\ref{fig9m}

At constant particle number and entropy, the volume of the trap will determine the temperature of the confined system, except in the small region where the temperature is not monotonic. 

To understand this behavior, we will consider the variation of the entropy per particle ($s = S/N$) under increase of particle number at constant temperature and volume. For $kT=0.3$, the reader can see from Fig.~\ref{fig9m} that starting from small particle numbers (large specific volumes), $s$ will decrease as the number of particles increases, once the horizontal line $k_BT = 0.3$ will cross curves with descending values of $s$. But around $v = 2L/N = 0.4$, there will be a small range of particle numbers where $s$ will increase. Afterwards, $s$ starts to decrease again, as illustrated in Fig.~\ref{fig10m} (insets of panels a, b and c). The derivative of the entropy per particle with respect to the particle number is given by
\beq
\frac{\partial}{\partial N}\left(\frac{S}{N}\right) = \frac{1}{N}\left(\stdif{S}{N} - \frac{S}{N}\right).
\eeq
A positive value for this derivative corresponds to a positive value for $\tdif{T}{v}{s}$,
\beq
\tdif{s}{N}{V,T} = \frac{vc_v}{NT}\tdif{T}{v}{s}.
\eeq
Therefore, a unusual positive derivative requires $\stdif{S}{N} > \frac{S}{N}$. We have decomposed the entropy of our system as the sum of the entropy of the non-interacting Kohn-Sham system plus the correlation entropy, $S = S_{KS} + S_c$.  The correlation entropy is computed approximately by LDA,
\beq
S_c = \sum_i s_c^\mathrm{hom}(n_i).
\eeq
At this point we need a close inspection of how the density profile changes as the particle number increases. This is illustrated in Fig.~\ref{fig10m}, which also displays the correlation entropy per particle of the homogeneous system as a function of density for $k_BT=0.3$.  At this low temperature, for $2L/N = 0.5$, the denstiy profile presents an almost flat central plateau with density equal to one. This is due to the energy gap in the charge sector of the one-dimensional Hubbard model. As the particle number increases, the plateau widens with essentially constant density as it can be seen in Fig.~\ref{fig10m}a.   

For larger particle numbers, a little bump appears above the central plateaus. The density in the middle of the trap is slightly above one. There is a signficant portion of the trap where the density is in the interval $[1,1.12]$, where the correlation entropy per particle grows very fast with the density (Fig.~\ref{fig10m}d). Within LDA, the contribution of the large derivatives $\stdif{s_c}{n}$ can eventually be enough to render $\stdif{S}{N} > \frac{S}{N}$. For still larger particle numbers, the density in the middle of the trap becomes higher than $1.12$ for which 
$\stdif{s_c}{n} < 0$ and the high density sites start to compensate the contribution from sites with $\stdif{s_c}{n} > 0$ and eventually we go back to the usual behavior $\stdif{S}{N} < \frac{S}{N}$ (which corresponds to $\tdif{T}{v}{s} < 0$). The important point is that the unusual increase of temperature under adiabatic expansion signals that the system will develop a density profile with a flat plateau in the center, which corresponds to an uncompressible region (the insulating phase of the Hubbard model). Starting with a given number of particles and a weakly confined system, we can get the insulating phase in the center of the trap by decreasing its volume only if the entropy per particle is low enough ($\lesssim 0.75$~$k_B$ from Fig.~\ref{fig9m}). For higher values of the entropy per particle, when the volume per particle is around 0.4, the temperature of the system will be already high enough to prevent the formation of a flat plateau in the density profile. From Fig.~\ref{fig9m}, for $s=0.8~k_B$ and $V/N = 0.4$ we have $k_BT = 0.57$. This energy should be compared with the energy gap in the charge sector of the model, $\Delta \approx 4.7$, to give $k_BT = 0.12\Delta$. Therefore, for temperatures larger than $0.1\Delta$, even with strong confinement, the insulating phase may not appear in the center of the trap. 

\section{Conclusion} 
\label{secconclusion}

In this work we have presented in detail a DFT approach to the thermodynamics of the inhomogeneous one-dimensional Hubbard model. The approach is based on the LDA for the correlation Helmholtz free-energy, where the data for the homogeneous model comes from numerical solution of QTM integral equations~\cite{klumper1996,klumper1998,essler2005}. The general formalism can be used with any external potential. Extensive comparison between TBALDA and exact diagonalization was done for a small system and TBALDA performed well, allowing its use to study the thermodynamics of larger inhomogeneous systems.

The thermodynamics for the fermionic one-dimensional Hubbard model confined by a parabolic external potential was studied. After introducing the system volume, the trap pressure was obtained, being equal to twice the external potential energy per unit of volume, and the thermodynamic description of the confined system was shown to proceed in the same way we classically describe the thermodynamics of a homogeneous system. One must note that the approach used here to treat the one-dimensional model could naturally be used to treat the Hubbard model in two and three dimensions. In theses cases, however, we would have a practical difficult to make computations based on DFT with LDA, because there is no exact solution for the homogeneous model as it happens in one dimension. Alternative considerations are required to build the energy functionals and this is ongoing research. It is also valid to note that other forms of external potential would lead to a similar thermodynamic description in terms of a volume appropriately defined. For example, with potentials of type $V(x) = |x/L|^p$, we could also define the volume by $2L$ and easily find the meaning of the trap pressure. 

The combination of reasonable accuracy and fast computation is the most appealing feature of a DFT approach. The procedure is not exact due to the necessary approximation to build the correlation energy functional, but the errors are typically small enough to let us use it with confidence. The DFT approach based on TBALDA allowed us to determine the thermodynamic properties of the harmonically confined Hubbard model. In this work we emphasized the behavior of the confined system at constant entropy due to the experimental motivation. We found an unusual increase of temperature under isentropic expansion, that can be of experimental interest as well as can motivate new calculations using more precise but much more demanding computational techniques, as QMC and exact diagonalization. The unusual increase of temperature was understood, at least within TBALDA, as a consequence of the peculiar density dependence of the correlation entropy per site for the homogeneous model around half-filling at low temperatures. Indication of how low the temperature and the entropy per particle must be to the insulating phase appear in the center of trap was given. This work opens the door to study, in the framework of DFT, the thermodynamics of the inhomogeneous fermionic one-dimensional Hubbard model in other physically interesting conditions, such as with negative $U$ and with spin polarization.

\section{Acknowledgments}  

The author is grateful to L. N. Oliveira for a careful reading of the manuscript
. Financial support from brazilian agency CNPq is acknowledged.


\begin{thebibliography}{99}

\bibitem{parrbook_kohnrev}
R. G. Parr and W. Yang, {\it Density-Functional Theory of Atoms and Molecules} (Oxford University Press, Oxford, 1989); W. Kohn, Rev. Mod. Phys. {\bf 71}, 1253 (1999).

\bibitem{hk1964}
P. Hohenberg and W. Kohn,
Phys. Rev. {\bf 136}, B864 (1964).

\bibitem{mermin1965}
N. D. Mermin,
Phys. Rev. {\bf 137}, A1441 (1965).

\bibitem{baroniRMP2001}
S. Baroni, S. de Gironcoli, A. Dal Corso, P. Giannozzi,
Rev. Mod. Phys. {\bf 73}, 515 (2001).

\bibitem{qha2010}
S. Baroni, P. Giannozzi, and E. Isaev,
Rev. Mineral. Geochem. {\bf 71}, 39 (2010).

\bibitem{payneRMP92}
M. C. Payne, M. P. Teter, D. C. Allan, T. A. Arias and J. D. Joannopoulos,
Rev. Mod. Phys. {\bf 64}, 1045 (1992).

\bibitem{tuckermanMDreview}
M. E. Tuckerman,
J. Phys.:Condens. Matter {\bf 14}, R1297 (2002).

\bibitem{abinitMDbook}
D. Marx and J. Hutter, {\it Ab Initio Molecular Dynamics: Basic Theory and Advanced Methods} (Cambridge University Press, New York, 2009).

\bibitem{renata2010}
R. M. Wentzcovitch, Y. G. Yu and Z. Wu,
Rev. Mineral. Geochem. {\bf 71}, 59 (2010).

\bibitem{desjarlais2002}
M. P. Desjarlais, J. D. Kress and L. A. Collins,
Phys. Rev. E {\bf 66}, 025401R (2002)

\bibitem{steelRev}
T Hickel, B. Grabowski, F. K\"{o}rmann and J. Neugebauer
J. Phys.:Condens. Matter {\bf 24}, 053202 (2012)

\bibitem{qmcT2013}
E. W. Brown, B. K. Clark, J. L. DuBois and D. M. Ceperley,
Phys. Rev. Lett. {\bf 110}, 146405 (2013).

\bibitem{fitqmcT2014}
V. V. Karasiev, T. Sjostrom, J. Dufty and S. B. Trickey,
Phys. Rev. Lett. {\bf 112}, 076403 (2014).

\bibitem{reviewCapelleCampo}
K. Capele and V. L. Campo,
Phys. Rep. {\bf 528}, 91 (2013).

\bibitem{kurth_prl_2011}
G. Stefanucci and S. Kurth,
Phys. Rev. Lett. {\bf 107}, 216401 (2011).

\bibitem{gao_temp13}
X. Gao, A. H. Chen, I. V. Tokatly and S. Kurth,
Phys. Rev. B {\bf 86}, 235139 (2012).

\bibitem{ibloch_rmp2008}
I. Bloch, J. Dalibard and W. Zwerger,
Rev. Mod. Phys. {\bf 80}, 885 (2008).

%\bibitem{rigol_ensemble}
%M. Rigol,
%Phys. Rev. A {\bf 72}, 063607 (2005).

\bibitem{takahashi1972}
M. Takahashi,
Prog. Theor. Phys. {\bf 47}, 69 (1972).

\bibitem{takahashibook}
M. Takahashi, {\it Thermodynamics of One-Dimensional Solvable Models} (Cambridge University Press, Cambridge, England, 1999).

\bibitem{takahashi2002}
M. Takahashi and M. Shiroishi,
 Phys. Rev. B {\bf 65}, 165104 (2002).

\bibitem{klumper1996} 
A. Kl\"{u}mper and R. Z. Bariev,
Nucl. Phys. B {\bf 458}, 623 (1996).

\bibitem{klumper1998} 
G. J\"{u}ttner, A. Kl\"{u}mper and J. Suzuki,
Nucl. Phys. B {\bf 522}, 471 (1998).

\bibitem{essler2005}
F. H. L. Essler, F. Holger, F. G\"{o}hmann, A. Kl\"{u}mper, and V. E. Korepin, {\sl The One-Dimensional Hubbard Model\/} (Cambridge University Press, Cambridge, 2005). 

\bibitem{rigol_dphases03}
M. Rigol, A. Muramatsu, G. G. Batrouni, R. T. Scalettar,
Phys. Rev. Lett. {\bf 91}, 130403 (2003)

\bibitem{drummond_dphases05}
X. -J. Liu, P. D. Drummond, H. Hu,
Phys. Rev. Lett {\bf 94}, 136406 (2005).

\bibitem{neemias2002}
 N. A. Lima, L. N. Oliveira and K. Capelle,
Europhys. Lett. {\bf 60}, 601 (2002).

\bibitem{neemias2003}
N. A. Lima, M. F. Silva, L. N. Oliveira and K. Capelle,
Phys. Rev. Lett. {\bf 90}, 146402 (2003). 

\bibitem{xianlong2006}
G. Xianlong, M. Polini, M. P. Tosi, V. L. Campo, K. Capelle, and M. Rigol,
Phys. Rev. B {\bf 73}, 165120 (2006). 

\bibitem{ks1965}
W. Kohn and L. J. Sham,
Phys. Rev. {\bf 140}, A1133 (1965).

\bibitem{gorelik2012} 
E. V. Gorelik, D. Rost, T. Paiva, R. Scalettar, A. Kl\"{u}mper and N. Bl\"{u}mer,
Phys. Rev. A {\bf 85}, 061602 (2012)

\bibitem{vito2012}
V. L. Campo, K. Capelle, C. Hooley, J. Quintanilla and V. W. Scarola,
Phys. Rev. A {\bf 85}, 033644 (2012).

%\bibitem{Ho_local}
%T.-L. Ho  and Q. Zhou, Nat. Phys. {\bf 6}, 131 (2010).

%\bibitem{shiba1972}
%H. Shiba and P. A. Pincus, Phys. Rev. {\bf 5}, 1966 (1972). 

\bibitem{Lima2007557}
N. A. Lima, A. L. Malvezzi and K. Capelle,
Solid State Commun., {\bf 144} 557 (2007).

\bibitem{physicaB}
V. L. Campo, J. Quintanilla and C. Hooley,
Physica B {\bf 404}, 3328 (2009).



\end{thebibliography}
\end{document}